\documentclass[prl,twocolumn,showpacs,reprint,superscriptaddress]{revtex4-1}

\usepackage{graphicx}
\usepackage{dcolumn,float}
\usepackage{amsmath,amssymb}
\usepackage{subfigure}
\usepackage{hyperref}
\usepackage{color}

\definecolor{venetianred}{rgb}{0.78,0.03,0.08}

\usepackage{etoolbox}

\newcommand{\fref}[1]{Fig.~\ref{#1}}
\newcommand{\eref}[1]{Eq.~(\ref{#1})}

\def\T{\mathrm{T}}
\def\id{id}
\def\J{\tilde{P}^{+}}
\def\Jf{\tilde{P}^{-}}
\def\Jbf{\tilde{P}^{\pm}}
\def\Jh{\hat{P}^{+}}
\def\Jhbf{\hat{P}^{\pm}}
\def\L{L^{+}}
\def\hF{K_0}
\def\Ktwo{K_2}
\def\Kone{K_1}
\def\kk{\kappa_1}
\def\v{\boldsymbol{v}}
\def\P{{\cal P}}
\def\K{{\cal K}}
\def\R{{\cal R}}
\def\A{\tilde{A}}
\def\rme{\mathrm{e}}
\def\rmd{\mathrm{d}}
\def\Wg{\mathrm{Wg^{O}}}

\begin{document}

\title{Multiparticle correlations in mesoscopic scattering: \\ boson sampling, birthday paradox, and Hong-Ou-Mandel profiles}

\newcommand{\RegensburgUniversity}{Institut f\"ur Theoretische Physik, 
Universit\"at Regensburg, D-93040 Regensburg, Germany}

\newcommand{\ShoUniversity}{Graduate School of Science and Engineering, Kagoshima University,
1-21-35, Korimoto, Kagoshima, Japan }

\author{Juan-Diego Urbina}
\affiliation{\RegensburgUniversity}
\author{Jack Kuipers}
\affiliation{\RegensburgUniversity}
\author{Sho Matsumoto}
\affiliation{\ShoUniversity}
\author{Quirin Hummel}
\affiliation{\RegensburgUniversity}
\author{Klaus Richter}
\affiliation{\RegensburgUniversity}

\begin{abstract}
The interplay between single-particle interference and quantum indistinguishability leads to signature correlations in many-body scattering. We uncover these with a semiclassical calculation of the transmission probabilities through mesoscopic cavities for systems of non-interacting particles. For chaotic cavities we provide the universal form of the first two moments of the transmission probabilities over ensembles of random unitary matrices, including weak localization and dephasing effects. If the incoming many-body state consists of two macroscopically occupied wavepackets, their time delay drives a quantum-classical transition along a boundary determined by the bosonic birthday paradox. Mesoscopic chaotic scattering of Bose-Einstein condensates is then a realistic candidate to build a boson sampler and to observe the macroscopic Hong-Ou-Mandel effect.                  
\end{abstract}

% \pacs{03.65.Sq, 05.45.Mt, 42.50.Ar, 73.23Ad}

\maketitle

In quantum mechanics  identical particles are indistinguishable and their very identity is then affected by quantum fluctuations and interference effects. A prominent type of Many-Body (MB) correlations is exemplified by the celebrated Hong-Ou-Mandel (HOM) effect \cite{HOM}, by now the standard indicator of MB coherence in quantum optics. There, the probability of observing two photons leaving in different arms of a beam splitter is measured. As a function of the delay between the arrival times of the incoming pulses, the coincidence probability shows a characteristic dip that can be seen as an effective Quantum-Classical Transition (QCT) where the difference in arrival times dephases the MB interference due to quantum indistinguishability \cite{QCT}. In recent years a wealth of hallmark experimental studies of MB scattering has gone beyond this scenario \cite{BSE0,BSE1,BSE2,HOMrec1,HOMrec2,BSE3,BSE4}. The aim is to reach a regime where for a random Single-Particle (SP) scattering matrix $\sigma$, and due to MB interference, the complexity in the calculation of MB scattering probabilities as a function of $\sigma$ beats classical computers, the Boson Sampling (BS) problem \cite{BSA}. However, while current optical devices \cite{BSE2,BSE4} reach photon occupations (below 6) far from the required regime of large number of particles, on platforms based on trapped ions \cite{BSTI}, cold atoms \cite{BSCA} and spin chains \cite{BSSC} it is not clear how to sample $\sigma$ uniformly.
\begin{figure}[ttt]
\includegraphics[width=0.92\columnwidth]{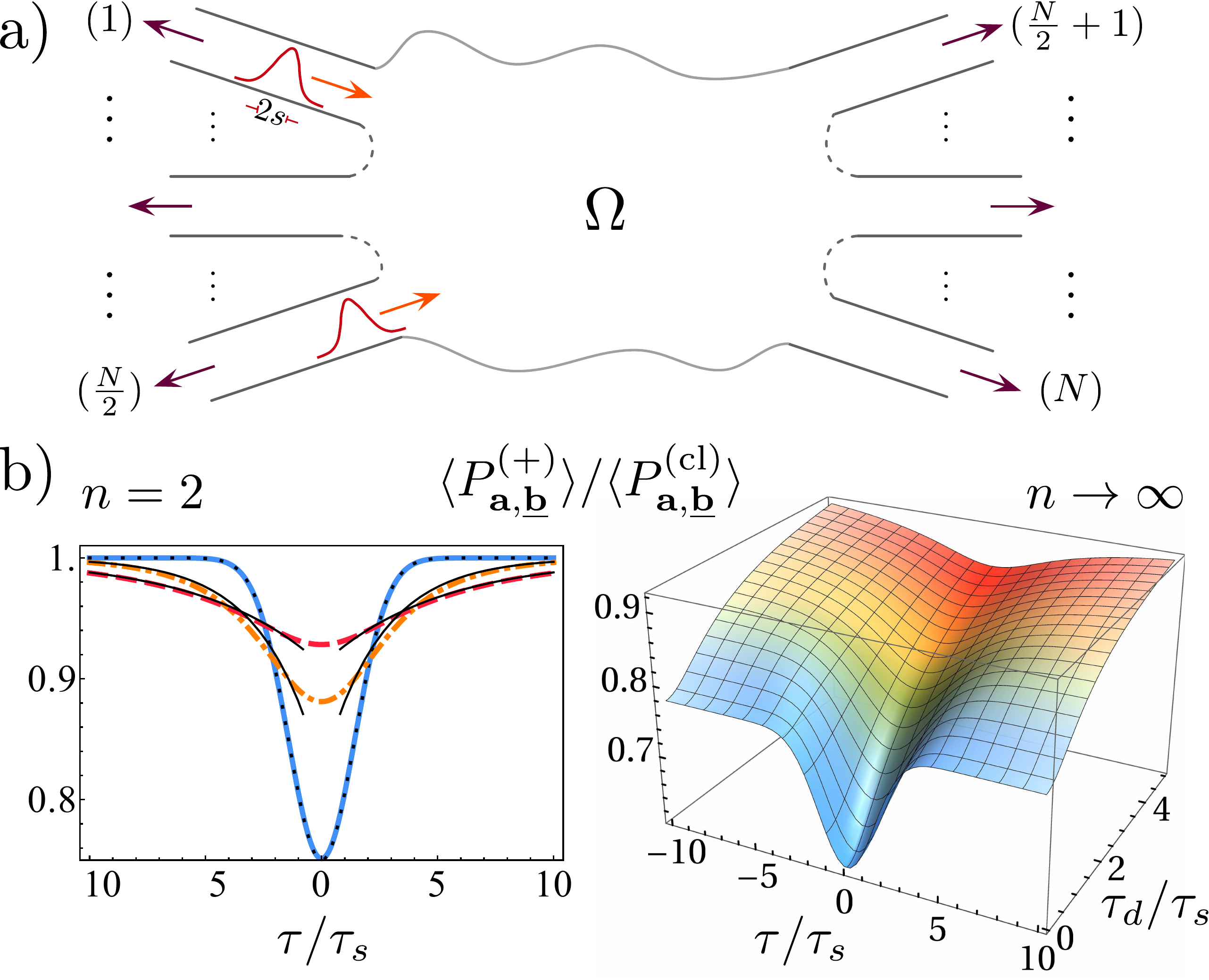}
\caption{a) Two bosonic wavepackets with mean velocity $v$, transversal channels ${\bf a}=(a_{1}, a_{2})$ and width $s=v\tau_{s}$ approach the chaotic cavity $\Omega$ with mean position difference $z=v\tau$. b) ratio $\langle P^{(+)}\rangle/\langle P^{({\rm cl})}\rangle$, between the quantum and classical probabilities (averaged over mesoscopic fluctuations) to find the bosonic particles in {\it different} output channels ${\bf \underline{b}}$. b) Left: For singly-occupied wavepackets, $n=2$ (with $N=4$ channels) we observe a generalized Hong-Ou-Mandel (HOM) profile that changes from Gaussian (dotted) to a universal exponential (thin solid tails) as function of the cavity's dwell time $\tau_{d}$, with $\tau_{d}/\tau_{s}=0.1, 2.5, 5$ (solid blue, dashed-dotted yellow, dashed red), Eqs.~(\ref{eq:P2}-\ref{eq:Q2lim}) with $z=z_{12}$. Right: For $n\to \infty,N=\alpha n^{\eta}$, $\langle P^{(+)}\rangle$ reaches its classical limit if $\eta >2$ or trivially saturates due to the Bosonic Birthday Paradox for $\eta < 2$. For $\eta=2$ the Quantum-Classical Transition shows an {\it exponentiated} HOM-like profile, Eq.~(\ref{eq:finI}) with $x=0,\alpha=1$.} 
\label{fig:SP1}
\end{figure}

Here we study {\it mesoscopic} MB scattering of {\it massive} particles depicted in Fig.~\ref{fig:SP1}(a). While formally identical to the optical situation in that it relates SP scattering matrices with MB scattering probabilities, it allows for large occupations through, e.g, Bose-Einstein condensation. Moreover, a standard result from Quantum Chaos \cite{Haa} says that complex SP interference due to classical chaos inside such a mesoscopic scattering cavity $\Omega$ transforms averages over small changes of the incoming energies into averages over an appropriate ensemble of unitary matrices, thus providing a genuine sampling over random scattering matrices. With experimental techniques for preparation of coherent macroscopic occupations \cite{BEC1}, chaotic scattering \cite{BEC2} and detection \cite{BEC3}, mesoscopic scattering of BECs contains all prerequisites of a realistic platform for BS, its certification \cite{us}, and related tasks \cite{new}. This is illustrated with the recent realization of the two-particle HOM effect using atomic beam splitters in \cite{BEC4}. 

\begin{figure*}
\centering
\includegraphics[width=0.9\textwidth, height=3.2cm]{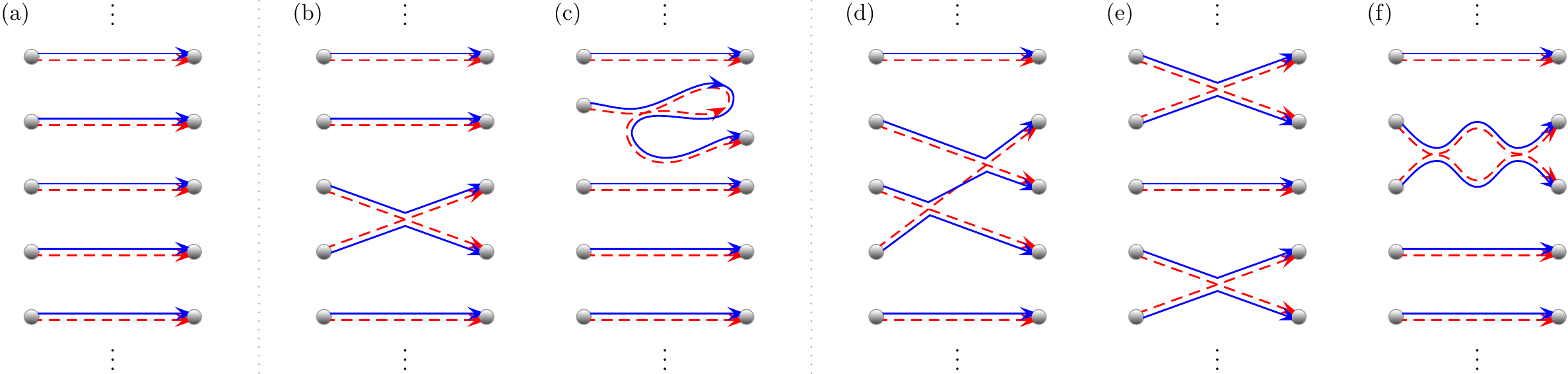}
\caption{Sets of interfering SP paths required for calculating MB transition probabilities, here for $n=5$. In (a), both SP and MB correlations are neglected. In (c), weak localization at the SP level is included. For (b,d,e), and (f) only MB correlations are included. Combined SP and MB effects appear when the links in a MB diagram are decorated with SP loops.}
\label{fig:jack}
\end{figure*}

Since the methods developed for the study of MB scattering of photons \cite{BST1,BST2,BST3,BST4,BST5} ignore mesoscopic effects and physical scales like the cavity's dwell time, we fill this gap and present analytic results on coherent MB scattering in the mesoscopic regime, particularly the way the QCT is affected by large occupation numbers and mesoscopic fluctuations. Supported by the universal correlations of SP scattering matrices \cite{BeeRMT,EQUI} responsible for characteristic mesoscopic wave interference effects like weak localization \cite{WLT} and universal conductance fluctuations \cite{UCFT}, we address the emergence of universal MB correlations due to the interplay between classical ergodicity, SP interference and quantum indistinguishability well beyond the standard semiclassical SP picture \cite{Brack}. Despite their intrinsically non-classical character, here MB correlations are successfully expressed and computed within a semiclassical approach in terms of interfering SP classical paths in the spirit of the Feynman path integral \cite{PathInt} by a one-to-one correspondence between MB classical paths (illustrated in Fig.~\ref{fig:jack}) and terms of the expansion of the MB scattering probabilities. Our complete enumeration and classification of the MB paths allows for an explicit analysis of emergent phenomena in the thermodynamic many-particle limit, something out of reach of leading-order Random Matrix Theory (RMT) methods \cite{BeeI,SkipI,SkipII}. 

We also show here how mesoscopic dephasing effects encoded in the dwell time lead eventually to a universal HOM profile, and  provide a mesoscopic approach to the Bosonic Birthday Paradox (BBP) that constrains the experimental realization of BS due to a counter-intuitive scaling of coincidence probabilities with the density of particles \cite{Arkh}. Our methods can be extended to the optical case by using the dispersion relation for photons and changing the cavity $\Omega$ to a multi-port waveguide network, making a connection with recent experiments \cite{HOMrec1,HOMrec2,BSE3,BSE4}. 

The set~up of the mesoscopic many-body scattering problem is depicted in Fig.~\ref{fig:SP1}(a). The incoming particles ($i\!=\!1,\ldots,n$) with positions $(x_{i},y_{i})$ occupy SP states represented by normalized wavepackets 
\begin{equation}
\phi_{i}(x_{i},y_{i})\propto {\rm e}^{-i k x_{i}}X(x_{i}-z_{i})\chi_{a_{i}}(y_{i}) \, .
\end{equation}
The longitudinal wavepackets ${\rm e}^{-i k x}X(x-z)$ have variance $s^{2}$, mean initial position $z \gg s$, and approach the cavity $\Omega$ with mean momentum $\hbar k=mv>0$ along the longitudinal directions $-x_{i}$. The relative positions of the incoming particles are then parametrized by the differences $z_{ij}\!=\!z_{i}-z_{j}$ or delay times $\tau_{ij}=z_{ij}/v$. The transverse wavefunction in the incoming channel $a_{i}\in\{1,\ldots,N/2\}$ is $\chi_{a_{i}}(y_{i})$ and has energy $E_{\rm \chi}$, assumed for simplicity to be identical for all channels. 

If the particles are identical, quantum indistinguishability demands their joint state to be symmetrized according to their spin \cite{sak}. Introducing $\epsilon=-1~(+1)$ for fermions (bosons), the symmetrized amplitude to find the particles leaving in channels ${\bf b}\!=\!(b_{1},\ldots,b_{n})$ with energies ${\bf E}\!=\!(E_{1},\ldots,E_{n})$ is given by a sum over the action of the $n!$ elements ${\cal P}$ of the permutation group,
\begin{equation}
\label{eq:Aeps}
A^{(\epsilon)}_{{\bf a},{\bf b}}({\bf E})=\sum_{{\cal P}}\epsilon^{{\cal P}}A_{{\bf a},{\cal P}{\bf b}}({\cal P}{\bf E}) \,  
\end{equation}
on the scattering amplitude for distinguishable particles,
\begin{equation}
\label{eq:Ampl}
A_{{\bf a},{\bf b}}({\bf E})\!=\!\prod_{i=1}^{n}\sqrt{\frac{m}{\hbar}}\frac{{\rm e}^{-i(k\!-\!q_{i})z_{i}}}{\sqrt{2\pi\hbar q_{i}}}\tilde{X}(k-q_{i})\sigma_{b_{i},a_{i}}(E_{i}) \, 
\end{equation}
where $\hbar q_{i}=\sqrt{2m(E_{i}-E_{\chi})}$ and $\tilde{X}(k)=\int{\rm e}^{-ikx}X(x)dx$. When $n=1$, Eq.~(\ref{eq:Ampl}) formally defines the SP scattering matrix $\sigma_{b,a}(E)$ connecting the incoming and outgoing channels $a$ and $b$.
With these definitions, the MB probability to find the particles leaving in channels ${\bf b}$ but regardless of their energies is given by
\begin{eqnarray}
\label{eq:MBP}
P^{(\epsilon)}_{{\bf a},{\bf b}}&=&\frac{1}{{\bf a}! {\bf b}!}\int_{E_{\chi}}^{\infty}d {\bf E} |A^{(\epsilon)}_{{\bf a},{\bf b}}({\bf E})|^{2} \\
&=&\frac{1}{{\bf a}! {\bf b}!}\sum_{{\cal P},{\cal P}'}\epsilon^{{\cal P}+{\cal P}'}\int_{E_{\chi}}^{\infty} d{\bf E}A_{{\bf a},{\cal P}{\bf b}}({\cal P}{\bf E}) A^{*}_{{\bf a},{\cal P}'{\bf b}}({\cal P}'{\bf E}). \nonumber
\end{eqnarray}
Equation~(\ref{eq:MBP}) includes the normalization factors ${\bf o}!=\prod_{i}{\rm mul}(o_{i})!$, where ${\rm mul}(o_{i})$ is the multiplicity of the channel index $o_{i}$, in order to have $\sum_{b_{1} \le \ldots \le b_{n}}^{n}P^{(\epsilon)}_{{\bf a},{\bf b}}=1$.

Due to interference between different (${\cal P} \ne {\cal P}'$) distinguishable MB configurations, $P^{(\epsilon)}_{{\bf a},{\bf b}}$ is sensitive to the relative positions of the incoming wavepackets $z_{ij}$. This dependence drives a transition from indistinguishability to effective distinguishability for $z_{ij}\to \infty$. MB interference due to indistinguishability is thus intrinsically dephased and one observes an effective QCT \cite{BSE3,BSE4}, as seen from the HOM scenario \cite{HOM} where $\sigma$ is ${\bf E}$-independent and $2n=4=N$. In this case we get, using Eq.~(\ref{eq:MBP}), 
\begin{equation}
\label{eq:HOMpure}
P^{{\rm HOM}}_{a_{1}\ne a_{2},b_{1} \ne b_{2}}=\frac{|[\sigma]|^{2}+1}{2}+\epsilon\frac{|[\sigma]|^{2}-1}{2}~{\cal F}^{2}(z_{12}) \, , 
\end{equation}
where $[\cdot]$ denotes permanent (unsigned determinant) and
\begin{equation}
\label{eq:over}
{\cal F}(z)=\int_{-\infty}^{\infty}X(x)X(x-z)dx \, , 
\end{equation}
satisfying ${\cal F}(0)=1, {\cal F}(\infty)=0$ is responsible for the non-universal profile of the QCT, as shown in the dotted curve in the left panel of Fig.~\ref{fig:SP1}(b) for Gaussian wavepackets. 

Individual $\sigma$-matrices with specific entries leading to Eq.~(\ref{eq:HOMpure}) and its few-particle generalizations are routinely constructed in arrays of beam splitters connecting waveguides for photonic systems \cite{BST2,BST3,HOMrec1,HOMrec2} and in quantum point contacts for electrons occupying edge states \cite{QPCE,QPCT}. Thanks to the Bohigas-Gianonni-Schmidt conjecture, replacing the beam splitter or point contact by a chaotic mesoscopic cavity allows to sample the moments $\langle f(\sigma)^{\mu} \rangle$ of any observable $f(\sigma)$ over the full ensemble of random, unitary matrices $\sigma$ by sampling over energy windows or small variations of the cavity \cite{Haa}. In this case averages of the form  $\langle \sigma_{b,a}(E) \sigma_{b',a'}^{*}(E')\rangle$ display universal features depending only on the presence or absence of time reversal invariance, denoted as the orthogonal ($\beta=1$) and unitary ($\beta=2$) case. Interference effects in SP scattering probabilities are semiclassically understood in terms of statistical correlations among classical actions \cite{EQUI,UCFT,WLT,BeeRMT,accorr} and here we generalize these methods.

We will mainly focus in the case, denoted by ${\bf \underline{b}}$, where every output channel is singly occupied; for $\beta=1$ we also demand that the in and outgoing channels are different. In our approach any $2n\mu$-order correlator of $\sigma$-matrices appearing in the moments $\langle |P^{(\epsilon)}_{{\bf a},{\bf \underline{b}}}({\bf E})|^{2\mu} \rangle$ of the distribution of scattering probabilities, Eq.~(\ref{eq:MBP}), is given by an infinite diagrammatic expansion with terms that can be visualized as a set of links joining $n\mu$ in and outgoing channels, see Fig.~\ref{fig:jack}. For the averaged transition probability, $\mu=1$, the classical limit  
\begin{equation}
\label{eq:PclassFA}
\langle P^{{\rm (cl)}}_{{\bf a},{\bf b}}\rangle=(n!/{\bf b}!)N^{-n} \, 
\end{equation}
for general ${\bf b}$, is obtained from the trivial topology in Fig.~\ref{fig:jack}(a) \cite{supp1}. In Eq.~(\ref{eq:PclassFA}) $N$ is the number of open channels at the mean initial SP energy $U=mv^{2}/2+E_{\chi}$. Quantum effects at the SP level, in the spirit of \cite{WLT,EQUI}, give the sole contribution for ${\cal P}={\cal P}'$ in Eq.~(\ref{eq:MBP}) and are generated by adding SP loops to the links, as in Fig.~\ref{fig:jack}(c). These terms, independent of $\epsilon$, can be evaluated up to infinite order to give (with $\langle P^{{\rm (cl)}}_{{\bf a},{\bf \underline{b}}}\rangle=n!N^{-n}$)
\begin{equation}
\label{eq:PSP}
\langle P^{({\rm SP})}_{{\bf a},{\bf \underline{b}}}\rangle=\langle P^{{\rm (cl)}}_{{\bf a},{\bf \underline{b}}}\rangle(1-(1-2/\beta)/N)^{-n}\, .
\end{equation}
To calculate $\langle P^{(\epsilon)}_{{\bf a},{\bf \underline{b}}}\rangle$ we must include genuine MB effects characterized by correlations between different SP paths, ${\cal P}\ne{\cal P}'$. The first MB diagrams without SP loops are depicted in Figs.~\ref{fig:jack}~(b),~(d),~(e) while Fig.~\ref{fig:jack}(f) shows the diagram~(b) with a loop between 2 particles.  The basic correlator in Fig.~\ref{fig:jack}(b) involving a single pair of correlated paths is \cite{UCFT,KS08}
\begin{eqnarray}
\label{eq:SPcorr2}
&&\langle\sigma_{b_{i},a_{i}}(E_{i}) \sigma_{b_{j},a_{j}}(E_{j})\sigma^{*}_{b_{i},a_{j}}(E_{j}) \sigma^{*}_{b_{j},a_{i}}(E_{i})\rangle {\rm \ \ \ \ }\\ && {\rm \ \ \ \ \ \ \ \ \ \ \ \ \ \ \ \ \ \ }= \frac{1}{N^{3}} \frac{\hbar^2}{\hbar^2+\tau_{d}^2(E_{i}-E_{j})^2} +{\cal O}\left(\frac{1}{N^{4}}\right),  \nonumber
\end{eqnarray}
where $\tau_{d}$ is the dwell time, the average time a particle with energy $(E_{i}+E_{j})/2$ remains within $\Omega$. Taking into account only pairs of correlated paths, Eq.~(\ref{eq:MBP}) gives \cite{supp2}
\begin{equation}
\label{eq:P2}
\frac{\langle P^{(\epsilon)}_{{\bf a},{\bf \underline{b}}}\rangle}{\langle P^{{\rm cl}}_{{\bf a},{\bf \underline{b}}}\rangle}=\frac{\langle P^{({\rm SP})}_{{\bf a},{\bf \underline{b}}}\rangle}{\langle P^{{\rm cl}}_{{\bf a},{\bf \underline{b}}}\rangle}-\frac{\epsilon}{N}\sum_{i<j}^{n}{\cal Q}^{(2)}(z_{ij}) +{\cal O}\left(\frac{1}{N^{2}}\right)\, ,
\end{equation}
with the generalized overlap integral Eq.~(\ref{eq:over}),
\begin{equation}
\label{eq:Q2}
{\cal Q}^{(2)}(z)=\int_{-\infty}^{\infty}{\cal F}^{2}(z-vt)\frac{{\rm e}^{-\frac{|t|}{\tau_{d}}}}{2\tau_{d}}dt \, . 
\end{equation}
In order to study the impact of mesoscopic effects in the HOM scenario we take $n=2$ and the sum in our Eq.~(\ref{eq:P2}) reduces to a single contribution with $i=1,j=2$. In the left pannel of Fig.~\ref{fig:SP1}(b) we plot $\langle P^{(\epsilon)}_{{\bf a},{\bf \underline{b}}}\rangle/\langle P^{{\rm cl}}_{{\bf a},{\bf \underline{b}}}\rangle$ as function of the mismatch distance $z=z_{12}$ between the incoming wavepackets in the case of broken time reversal invariance where Eq.~(\ref{eq:PSP}) gives $\langle P^{({\rm SP})}_{{\bf a},{\bf \underline{b}}}\rangle=\langle P^{{\rm (cl)}}_{{\bf a},{\bf \underline{b}}}\rangle$. We see how mesoscopic effects produce universal deviations from the usual Gaussian profile, represented by the dotted line. 

The functions ${\cal Q}^{(2)}$ determine how the mismatch of arrival times dephases the MB correlations. We interpret Eqs.~(\ref{eq:P2},\ref{eq:Q2}) as follows: Pairs of incoming particles that are effectively distinguishable get to interfere if their time delay $\tau_{ij}$ in entering the cavity is compensated by the time $\tau_{d}$ the first particle is held within the mesoscopic scattering region. However, the interference gets weighted by the survival probability ${\rm e}^{-t/\tau_{d}}/\tau_{d}$ of remaining inside the chaotic scatterer $\Omega$. Universality of the dephasing of MB correlations is expected if $\tau_{d}$ competes with the delay times $\tau_{ij}$ and widths $\tau_{s}=s/v$ of the incoming wavepackets, and leads to exponential tails in the interference profile for $|z_{ij}| \gg s$. As shown in the left panel of Fig.~\ref{fig:SP1}(b) these exponential regions grow with the ratio $\tau_{d}/\tau_{s}$, while for $\tau_{d} \le \tau_{s}$ QCT depends on the shape of the incoming wavepackets, as in Eq.~(\ref{eq:HOMpure}), 
\begin{equation}
\label{eq:Q2lim}
{\cal Q}^{(2)}(z)\left\{\begin{array}{ll}\xrightarrow{v\tau_{d}\gg v\tau_{s}>k^{-1}} & \left(\int_{-\infty}^{\infty}{\cal F}^{2}(z)\frac{dz}{s}\right)\frac{{\rm e}^{-\frac{|z|}{v\tau_{d}}}}{2\tau_{d}/\tau_{s}} \\
\xrightarrow{v\tau_{s} \gg v\tau_{d}>k^{-1}} & {\cal F}^{2}(z). 
\end{array} \right. \, 
\end{equation}

Mesoscopic dephasing of two-particle interference plays a fundamental role in the thermodynamic limit $N,n\to \infty$ of the QCT through the mesoscopic version of the BBP \cite{Arkh} which constrains the scaling $N=\alpha n^{\eta}$ in a way that $\langle P^{(\epsilon)}_{{\bf a},{\bf b}}\rangle$ does not get trivially saturated either classically or by quantum bunching and antibunching \cite{BSA,BSE0,Arkh,MalNew,Shch}. To achieve a semiclassical theory of the mesoscopic BBP, in \cite{supp3} we use RMT techniques to calculate $\langle P^{(\epsilon)}_{{\bf a},{\bf b}}\rangle$, which is only possible for $z_{ij}=0, \tau_{d}/\tau_{s}=0$. We obtain the expression, valid for arbitrary $\epsilon,N,n,{\bf a},{\bf b}$ if $\beta=2$ and with the only condition ${\bf a} \cap {\bf b}=\varnothing$ if $\beta=1$,  
\begin{equation}
\label{eq:Pconj}
\left.\langle P^{(\epsilon)}_{{\bf a},{\bf b}}\rangle\right|_{z_{ij}=0}^{\frac{\tau_{d}}{\tau_{s}}=0}= \frac{{\cal W}^{(\epsilon)}_{\beta}(N,n)n!}{\prod_{l=0}^{n-1}(N+\epsilon l)} (\delta_{\epsilon,+}+\delta_{\epsilon,-}\delta_{{\bf b},{\bf \underline{b}}})\, ,
\end{equation}
with $\delta_{{\bf b},{\bf \underline{b}}}=1(0)$ if ${\bf b}$ is (is not) singly-occupied and
\begin{equation}
{\cal W}^{(\epsilon)}_{1}(N,n)=\frac{N+\epsilon(n-1)}{N+n+\epsilon(n-1)}{\rm \ , \ }{\cal W}^{(\epsilon)}_{2}(N,n)=1.
\end{equation}
Equation~(\ref{eq:Pconj}) is a generalization for arbitrary $\beta$ and $\epsilon$ of the bosonic, unitary case reported in \cite{Arkh}. A key observation is that, contrary to the  distinguishable (classical) case, Eq.~(\ref{eq:PclassFA}), result~(\ref{eq:Pconj}) is constant over the MB final states for $\beta=2$. SP chaos leads then to full MB equilibration for systems with broken time-reversal symmetry, providing dynamical support to the analysis of \cite{Arkh}.  

For singly-occupied states ${\bf \underline{b}}$, Eqs.~(\ref{eq:PclassFA},\ref{eq:Pconj}) give 
\begin{equation}
\label{eq:finII}
\left.\left(\frac{\langle P^{(\epsilon)}_{{\bf a},{\bf \underline{b}}}\rangle}{\langle P^{{\rm (cl)}}_{{\bf a},{\bf \underline{b}}}\rangle}\right)^{\epsilon}\right|_{z_{ij}=0}^{\frac{\tau_{d}}{\tau_{s}}=0}\xrightarrow[N=\alpha n^{\eta}]{n\gg1}\left\{\begin{array}{ll}
0 & {\rm for \ \ }\eta<2 \\
{\rm e}^{-\frac{1}{2\alpha}} & {\rm for \ \ }\eta=2 \\
1 & {\rm for \ \ }\eta>2 \, ,
\end{array}\right. 
\end{equation}
showing how in the thermodynamic limit scattering of identical particles is classical in the dilute limit $\eta>2$, it gets saturated due to boson bunching and fermion antibunching even at zero densities if $\eta<2$, and only the scaling $N=\alpha n^{2}$ gives a non-trivial limit. This is the essence of the BBP \cite{Arkh,BSA,MalNew,Shch}, here derived from RMT (and for $\eta>1$ from semiclassics) arguments for arbitrary $\beta,\epsilon$. For $\beta=1$, weak localization corrections to MB equilibration (akin to MB coherent backscattering \cite{Tom}), to BS and to BBP are obtained from Eq.~(\ref{eq:Pconj}). 

To address the interplay between intrinsic ($z_{ij}\ne 0$) and mesoscopic ($\tau_{d}/\tau_{s} \ne 0$) dephasing one must go beyond RMT and we resort to semiclassical diagrammatics. In \cite{supp4} we study the semiclassical generating function for $\langle P^{(\epsilon)}_{{\bf a},{\bf b}}\rangle$ and show that, order by order in the $1/N$ expansion, diagrams with pairwise correlations between particles like Fig.~\ref{fig:jack}(b),(e) dominate the $n\to \infty$ limit leading to Eq.~(\ref{eq:finII}) for $\eta>1$. The whole set of semiclasssical diagrams with pairwise correlations can now be constructed for $\tau_{d}/\tau_{s} > 0$ and $z_{ij} \ne 0$, and resumed to infinite order where the scaling $\eta=2$ emerges \cite{supp5}.   

If $z_{ij}\in\{0,z\}$, a situation that can be realized for bosons by injecting two wavepackets with macroscopic occupations $n(1\pm x)/2$, we get \cite{supp6}
\begin{equation}
\label{eq:finI}
\frac{\langle P^{(\epsilon)}_{{\bf a},{\bf \underline{b}}}\rangle}{\langle P^{{\rm (cl)}}_{{\bf a},{\bf \underline{b}}}\rangle}\xrightarrow[{N=\alpha n^{2}}]{n\gg 1}{\rm e}^{-\frac{\epsilon}{4\alpha}\left[\left(1+x^{2}\right){\cal Q}^{(2)}(0)+\left(1-x^{2}\right){\cal Q}^{(2)}(z)\right]} \, .
\end{equation} 
Remarkably then, for macroscopically populated incoming states we observe again a QCT driven by the arrival difference, with an {\it exponentiated} HOM-like profile, as shown in Fig.~\ref{fig:SP1}(b, right) for $x=0$ and $\alpha=1$.

\begin{figure}[ttt]
\includegraphics[width=8cm]{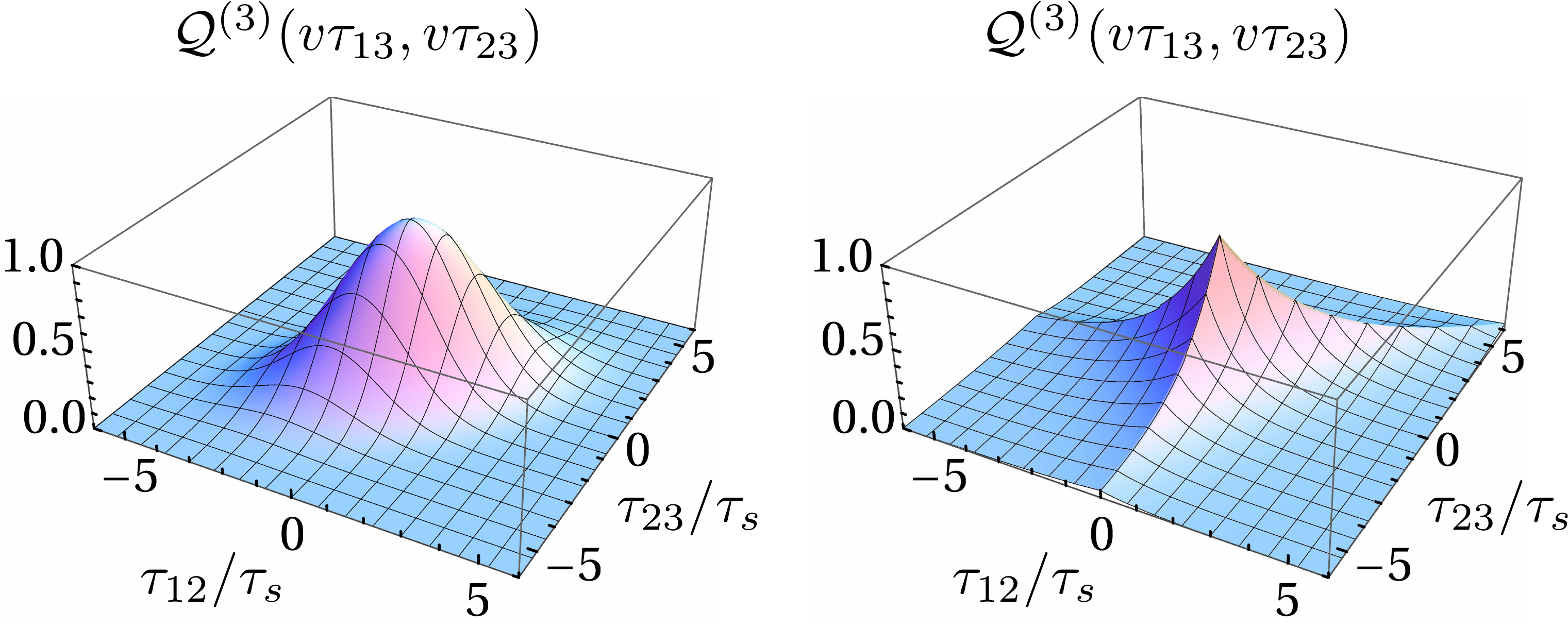}
\caption{Transition between the overlapping ($\tau_{d}/\tau_{s}=0.1$, left) and the universal exponential ($\tau_{d}/\tau_{s}=2$, right) regime for the three-body interference term, Eq.~(\ref{eq:P3}).} 
\label{fig:jd}
\end{figure}
Coming back to finite systems where MB interference is affected by other types of correlations, the diagram Fig.~\ref{fig:jack}(d) containing three-body correlations gives 
\begin{equation}
\label{eq:P3}
\frac{\langle P^{(\epsilon)}_{{\bf a},{\bf \underline{b}}}\rangle^{{\rm triplets}}}{\langle P^{{\rm (cl)}}_{{\bf a},{\bf \underline{b}}}\rangle}=\frac{2\epsilon}{N^{2}}\sum_{i<j<k}{\cal Q}^{(3)}(z_{ij},z_{kj}) \, , 
\end{equation}
with overlapping and exponential regimes given by  
\begin{equation}
{\cal Q}^{(3)}(z,z')\left\{\begin{array}{ll}\xrightarrow{v\tau_{d}\gg v\tau_{s} > k^{-1}} & {\cal C}^{(3)}\frac{{\rm e}^{-\frac{3{\rm Max}(z,z',0)}{v\tau_{d}}}}{2\tau_{d}/\tau_{s}} \frac{{\rm e}^{\frac{z+z'}{v\tau_{d}}}}{2\tau_{d}/\tau_{s}}, \\
\xrightarrow{v\tau_{s} \gg v\tau_{d} > k^{-1}} & {\cal F}(z) {\cal F}(z') {\cal F}(z-z') 
\end{array} \right. \, 
\end{equation}
and ${\cal C}^{(3)}=s^{-2}\int_{-\infty}^{\infty}{\cal F}(z) {\cal F}(z') {\cal F}(z-z')dzdz'$. As shown in Fig.~\ref{fig:jd}, this transition produces universal dephasing characterized by kinks with three-fold symmetry as a function of the time delay between incoming particles, consistent with the correlations measured in \cite{BSE4}. 

In conclusion, we have presented a semiclassical approach to quantum scattering for Many-Body systems and used it to study the emergence of universal effects due to the interplay between Single-Particle classical chaos and quantum correlations coming from indistinguishability. We have explicitly constructed the correlations responsible of Many-Body interference in mesoscopic scattering and computed their effect for both small and macroscopically large occupations in the thermodynamic limit, thus opening the possibility of translating Boson Sampling, the Bosonic Birthday Paradox and related timely problems into experimentally accessible scenarios of chaotic scattering with massive particles such as cold atoms, as outlined in the introduction. Single-Particle chaos turns out to be sufficient to achieve Many-Body ergodicity, and this allows us to compute mesoscopic corrections to the Bosonic Birthday Paradox. It leads to a sharp Quantum-Classical Transition in the thermodynamic limit and, under the scaling for the Quantum-Classical boundary, we found an exponentiated form of the Hong-Ou-Mandel profile. 

Going beyond the first moment $\langle P\rangle$ of the distribution of scattering probabilities, in \cite{supp7} we further calculate the leading order of the second moment $\langle P^{2}\rangle$. In fact, determining just the leading order of higher moments should be pertinent for the {\it Permanent Anti-Concentration Conjecture} important for Boson Sampling \cite{BSA}. Intriguingly then, semiclassical diagrams and random matrices open up new avenues for understanding permanent statistics, while mesoscopic scattering of massive bosons appears as a promising candidate for their measurement.  

We thank Andreas Buchleitner and Malte Tichy for instructive discussions, and an anonymous referee for valuable suggestions and for drawing our attention to references \cite{new1,new2}.

\cleardoublepage

\newcounter{sectionSupp} \setcounter{sectionSupp}{0}
\newcounter{subsectionSupp}[sectionSupp] \setcounter{subsectionSupp}{0}
\newcounter{subsubsectionSupp}[subsectionSupp] \setcounter{subsubsectionSupp}{0}

\renewcommand{\thesectionSupp}{\Roman{sectionSupp}}

\newcommand{\sectionSupp}[1]{%
	\section{}%
	\refstepcounter{sectionSupp}%
	\vspace*{-10mm}%
	\begin{center}{\small\bf\Roman{sectionSupp}.\hspace*{1em}\MakeTextUppercase{#1}}\end{center}%
	\vspace*{1.4mm}%
}

\newcommand{\subsectionSupp}[1]{%
	\subsection{}%
	\stepcounter{subsectionSupp}%
	\vspace*{-10mm}\begin{center}{\small\bf\Alph{subsectionSupp}.\hspace*{1em}#1}\end{center}%
}

\newcommand{\subsubsectionSupp}[1]{%\needspace{5\baselineskip}%
	\subsubsection{}%
	\stepcounter{subsubsectionSupp}%
	\vspace*{-8mm}%
	\begin{center}{\small\it\arabic{subsubsectionSupp}.\hspace*{1em}#1}\end{center}%
	\vspace*{1mm}%
}

\appendix

\onecolumngrid
\begin{center}
	\textbf{\Large{Supplementary material to the paper \\ "Multiparticle correlations in complex scattering: \\ birthday paradox and Hong-Ou-Mandel profiles in mesoscopic systems"}}
\end{center}
\vspace*{\columnsep}
\twocolumngrid

\section*{Introduction}

The statistical study of quantum correlations due to indistinguishability in MB mesoscopic scattering can be carried out in two different, complementary ways. The powerful random matrix theory techniques introduced in Sec.~\ref{sec:SuppJII} are suitable to address the universal regime where Hong-Ou-Mandel (HOM) effects can be neglected, namely, when the incoming wavepackets are mathematically taken as plane waves without a well defined position. Within random matrix theory, therefore, the effect of finite dwell time in the HOM profile is by definition irrelevant. In order to study the emergence of universality due to chaotic scattering on HOM profiles and the quantum-classical transition due to effective distinguishability, in Sec.~\ref{sec:SuppJI} a semiclassical theory is implemented. The semiclassical approach is able to account for the effect of localized, shifted incoming wavepackets in the universal limit where only pairwise correlations are relevant, as shown in Sec.~\ref{sec:SuppPair}. 
Before that, in Sec,~\ref{sec:SuppJD} the profile of the mesoscopic HOM effect is explicitly calculated.

\sectionSupp{Calculation of the generalized overlap integrals ${\cal Q}^{2}(z)$} 
\label{sec:SuppJD}

Using the definition Eq.~(\ref{eq:MBP}), the amplitudes given in Eq.(\ref{eq:Ampl}) and the correlator in Eq.~(\ref{eq:SPcorr2}) of the main text, we get
\begin{eqnarray}
\label{eq:Pintex}
{\cal Q}^{(2)}(z)&=&\int_{E_{\chi}}^{\infty}dE_{1}dE_{2} \frac{{\rm e}^{i(q_{2}-q_{1})z}}{1+\left[ \frac{\tau_{d}(E_{1}-E_{2})}{\hbar}\right]^{2}} \\&\times&\frac{m^{2}}{4\pi^{2}\hbar^{2}}\frac{|\tilde{X}(k-q_{1})|^{2}|\tilde{X}(k-q_{2})|^{2}}{\hbar^{2} q_{1}q_{2}} \, .  \nonumber
\end{eqnarray}
To further proceed, we use $E_{i}=E_{\chi}+\hbar^{2}q_{i}^{2}/2m$ and $q=q_{2}-q_{1},Q=(q_{1}+q_{2})/2$. Then we observe that in the momentum representation the incoming wavepackets $\tilde{X}(q_{i}-k)$ are strongly localized around $q_{1}=q_{2}=k$. As long as $ks \gg 1$ we can extend the lower limit of the integrals to $-\infty$ and keep only terms of first order in $q$. Under these conditions Eq.~(\ref{eq:Pintex}) yields
\begin{eqnarray}
{\cal Q}^{(2)}(z)&=&\int_{-\infty}^{\infty}dQdq\frac{{\rm e}^{iqz}}{1+v^{2}\tau_{d}^{2}q^{2}}  \\ &\times&\frac{|\tilde{X}(Q-k-q/2)|^{2} |\tilde{X}(Q-k+q/2)|^{2}}{4\pi^{2}} \, , \nonumber
\end{eqnarray}
which can be finally transformed into
\begin{equation}
\label{eq:PintexFin}
{\cal Q}^{(2)}(z)=\int_{-\infty}^{\infty}{\cal F}^{2}(z-vt)\frac{{\rm e}^{-\frac{|t|}{\tau_{d}}}}{2\tau_{d}}dt \, . 
\end{equation}

\sectionSupp{RMT approach for transmission probabilities}
\label{sec:SuppJII}

In this section we derive and then prove results for the transmission probabilities of bosons and fermions in both symmetry classes, Eq.~(\ref{eq:Pconj}) in the main text.

\subsectionSupp{Bosons}

For $n$ bosons, we start with the expression
\begin{equation} \label{Adef}
\A^{+}_n=\frac{1}{\sqrt{n!}} \sum_{\P\in S_n} \prod_{k=1}^n Z_{i_k,o_{\P(k)}} \, ,
\end{equation}
where we sum over all permutations $\P$ of $\{1,\ldots,n\}$ and where $Z=\sigma^{\T}$ is the transpose of the single particle scattering matrix $\sigma$ (so we can identify the first subscript as an incoming channel and the second as an outgoing one).  For simplicity we assume that all the channels are distinct.  This quantity is related to the $n$-particle amplitude when the particle energies coincide or $\tau_{d}=0$.

For the transmission probability we are interested in 
\begin{equation} \label{Asquareddef}
\vert \A^{+}_n\vert^2 = \A^{+}_n (\A^{+}_n)^{*} = \frac{1}{n!} \sum_{\P,\P'\in S_n} \prod_{k=1}^n Z_{i_k,o_{\P(k)}}Z^{*}_{i_{k},o_{\P'(k)}} \, .
\end{equation}
The averages over scattering matrix elements are known both semiclassically and from RMT (see \cite{bb96,bk13a} for example)
\begin{eqnarray} \label{scatcorrel}
& & \left \langle Z_{a_1,b_{1}} \cdots Z_{a_n,b_{n}}Z^{*}_{\alpha_{1},\beta_{1}} \cdots Z^{*}_{\alpha_{n},\beta_{n}} \right \rangle \\ 
& = & \sum_{\sigma,\pi \in S_n} V_{N}(\sigma^{-1}\pi) \prod_{k=1}^{n}\delta(a_{k}-\alpha_{\sigma(k)})\delta(b_{\bar{k}}-\beta_{\bar{\pi(k)}}) \, , \nonumber
\end{eqnarray}
where $V$ are class coefficients which can be calculated recursively.

However, since the channels are distinct, for each pair of permutations $\P,\P'$ in \eref{Asquareddef} only the term with $\sigma=\id$ and $\pi =\P (\P')^{-1}$ in \eref{scatcorrel} contributes.  One then obtains the result
\begin{equation} \label{Asquaresimp}
\J_n = \langle \vert \A^{+}_n\vert^2 \rangle = \frac{1}{n!} \sum_{\P,\P'\in S_n} V_{N}(\tau) \, ,
\end{equation}
where
\begin{equation}
\J_n =\frac{1}{n!}\left.\langle P^{+}_{{\bf a},{\bf b}}\rangle\right|_{z_{ij}=0}^{\frac{\tau_{d}}{\tau_{s}}=0}   \, 
\end{equation}
is the transmission probability when the particles enter at equal energies at the same time.  In \eref{Asquaresimp} $\tau=\P (\P')^{-1}$ is the target permutation of the scattering matrix correlator.  Since $\tau$ is a product of two permutations, summing over the pair $\P,\P'$ just means that $\tau$ covers the space of permutations $n!$ times and
\begin{equation} \label{Asquaresimpler}
\J_n = \sum_{\tau \in S_n} V_{N}(\tau) \, .
\end{equation}
Since the class coefficients only depend on the cycle type of $\tau$, one could rewrite the sum in terms of partitions.  For this we let $\v$ be a vector whose elements $v_l$ count the number of cycles of length $l$ in $\tau$ so that $\sum_l lv_l = n$.  Accounting for the number of ways to arrange the $n$ elements in cycles, one can write the correlator as
\begin{equation} \label{Asquarepart}
\J_n = \sum_{\v}^{\sum_l lv_l = n} \frac{n!V_{N}(\v)}{\prod_{l} l^{v_l} v_l!} \, ,
\end{equation}
where we represent the argument of $V$ by the cycles encoded in $\v$.

Typically, one considers correlators with a fixed target permutation, rather than sums over correlators as here in \eref{Asquaresimpler}.  For example fixing $\tau = (1,\ldots,n)$ provides the linear transport moments while $\tau=\id$ gives the moments of the conductance.  A summary of some of the transport quantities which have been treated with RMT and semiclassics can be found in \cite{bk13b}.

\subsubsectionSupp{Examples}

Representing the argument of the class coefficients $V_N$ instead by its cycle type, one can directly write down the result for $n=1,2$:  
\begin{eqnarray} \nonumber
\J_1 &=& V_N(1) \\
\J_2 &=& V_N(1,1) + V_N(2) \, ,
\end{eqnarray}
while for $n=3$ there are 6 permutations 
\begin{eqnarray} \nonumber \label{nis3perms}
& & (1)(2)(3) \qquad (123) \qquad (132) \\
& & (1)(23) \qquad (12)(3) \qquad (13)(2) \, ,
\end{eqnarray}
and so
\begin{equation}
\J_3 = V_{N}(1,1,1) + 3V_{N}(2,1) + 2V_N(3) \, .
\end{equation}

With the recursive results in \cite{bb96,bk13a} for the class coefficients we can easily find the following results for low $n$:

\pagebreak
\subsubsectionSupp{Unitary case}

Without time reversal symmetry, the results are
\begin{eqnarray} \label{Jresultsunit}
\J_1 &=& \frac{1}{N} \nonumber \\
\J_2 &=& \frac{1}{N(N+1)} \nonumber \\
\J_3 &=& \frac{1}{N(N+1)(N+2)} \nonumber \\
\J_4 &=& \frac{1}{N(N+1)(N+2)(N+3)} \nonumber \\
\J_5 &=& \frac{1}{N(N+1)(N+2)(N+3)(N+4)} \, .
\end{eqnarray}
The pattern
\begin{equation} \label{Jnallnunit}
\J_n = \frac{\Gamma(N)}{\Gamma(N+n)} \, .
\end{equation}
holds for all $n$ as we prove in a following subsection.  In fact we can relate $n! \J_n$ to the moments of a single element of a CUE random matrix and find a proof of \eref{Jnallnunit} in \cite{novak07}.

For a comparison to diagrammatic results, the expansion in $N^{-1}$ is
\begin{equation} \label{Jnunitexpand}
\J_n = \frac{1}{N^{n}} - \frac{n(n-1)}{2N^{n+1}} +\ldots
\end{equation}

\subsubsectionSupp{Orthogonal case}

With time reversal symmetry, the results are
\begin{eqnarray} \label{Jresultsorth}
\J_1 &=& \frac{1}{(N+1)} \nonumber \\
\J_2 &=& \frac{1}{N(N+3)} \nonumber \\
\J_3 &=& \frac{1}{N(N+1)(N+5)} \nonumber \\
\J_4 &=& \frac{1}{N(N+1)(N+2)(N+7)} \nonumber \\
\J_5 &=& \frac{1}{N(N+1)(N+2)(N+3)(N+9)} \, ,
\end{eqnarray}
with a general result of
\begin{equation} \label{Jnallnorth}
\J_n = \frac{\Gamma(N)}{\Gamma(N+n)}\frac{(N+n-1)}{(N+2n-1)} \, ,
\end{equation}
and an expansion of
\begin{equation} \label{Jnorthexpand}
\J_n = \frac{1}{N^{n}} - \frac{n(n+1)}{2N^{n+1}} +\ldots .
\end{equation}
For a proof of \eref{Jnallnorth} we can show that $n!\J_n$ coincides exactly with the moments of a single element of a COE random matrix.  The result as proved in \cite{matsumoto12} leads directly to \eref{Jnallnorth}.

\subsectionSupp{Fermions}

For $n$ fermions we start instead with
\begin{equation} \label{Bdef}
\A^{-}_n=\frac{1}{\sqrt{n!}} \sum_{\P\in S_n} (-1)^\P\prod_{k=1}^n Z_{i_k,o_{\P(k)}} \, ,
\end{equation}
where $(-1)^{\P}$ represents the sign of the permutation, counting a factor of -1 for each even length cycle in $\P$. Following the same steps for bosons, one has
\begin{equation} \label{Bsquaresimp}
\Jf_n = \langle \vert \A^{-}_n \vert^{2} \rangle = \sum_{\tau \in S_n} (-1)^{\tau} V_{N}(\tau) \, ,
\end{equation}
so for example
\begin{equation}
\Jf_3 = V_{N}(1,1,1) - 3V_{N}(2,1) + 2V_N(3) \, .
\end{equation}

Calculating the class coefficients recursively one then finds the following results for low $n$:

\subsubsectionSupp{Unitary case}

Without time reversal symmetry, the results are
\begin{eqnarray} \label{Jfresultsunit}
\Jf_1 &=& \frac{1}{N} \nonumber \\
\Jf_2 &=& \frac{1}{N(N-1)} \nonumber \\
\Jf_3 &=& \frac{1}{N(N-1)(N-2)} \nonumber \\
\Jf_4 &=& \frac{1}{N(N-1)(N-2)(N-3)} \nonumber \\
\Jf_5 &=& \frac{1}{N(N-1)(N-2)(N-3)(N-4)} \, ,
\end{eqnarray}
The pattern turns out to be
\begin{equation} \label{Jfnallnunit}
\Jf_n = \frac{\Gamma(N-n+1)}{\Gamma(N+1)} \, ,
\end{equation}
and the expansion in $N^{-1}$ is
\begin{equation} \label{Jfnunitexpand}
\Jf_n = \frac{1}{N^{n}} + \frac{n(n-1)}{2N^{n+1}} +\ldots
\end{equation}

\subsubsectionSupp{Orthogonal case}

With time reversal symmetry, the results are
\begin{eqnarray} \label{Jfresultsorth}
\Jf_1 &=& \frac{1}{(N+1)} \nonumber \\
\Jf_2 &=& \frac{1}{(N+1)N} \nonumber \\
\Jf_3 &=& \frac{1}{(N+1)N(N-1)} \nonumber \\
\Jf_4 &=& \frac{1}{(N+1)N(N-1)(N-2)} \nonumber \\
\Jf_5 &=& \frac{1}{(N+1)N(N-1)(N-2)(N-3)} 
\end{eqnarray}
with a general result of
\begin{equation} \label{Jfnallnorth}
\Jf_n = \frac{\Gamma(N-n+2)}{\Gamma(N+2)}
\end{equation}
and an expansion of
\begin{equation} \label{Jfnorthexpand}
\J_n = \frac{1}{N^{n}} + \frac{n(n-3)}{2N^{n+1}} +\ldots
\end{equation}

\subsectionSupp{Proofs}

Now we can turn to proving the formulae in \eref{Jfnallnunit} and \eref{Jfnallnorth}.  These proofs build heavily on \cite{matsumoto12, matsumoto13} for the underlying details and methods.  To introduce the techniques though, we start with the simpler case of reproving \eref{Jnallnunit}.

\subsubsectionSupp{Unitary bosons}

Starting with the sum over permutations in \eref{Asquaresimpler}, we use the fact that the class coefficients, which are also known as the unitary Weingarten functions admit the following expansion \cite{matsumoto13}
\begin{equation} \label{Jnunitproofbegin}
\J_n = \sum_{\tau \in S_n} V_{N}(\tau) = \frac{1}{n!} \sum_{\lambda \vdash n} \frac{f^{\lambda}}{C_\lambda(N)} \sum_{\sigma\in S_n} \chi^{\lambda}(\sigma) \, ,
\end{equation}
where $\lambda$ is a partition of $n$ and the remaining term are as defined in \cite{matsumoto13}.  To evaluate the sum we employ the character theory for symmetric groups.  The trivial character for $S_n$ is $\chi^{(n)}(\sigma)=1$ for $\sigma\in S_n$ while the orthogonality of irreducible characters means that
\begin{equation}
\frac{1}{n!}\sum_{\sigma\in S_n} \chi^{\lambda}(\sigma)\chi^{\mu}(\sigma) = \delta_{\lambda,\mu} \, .
\end{equation}
Combining both these facts we have
\begin{equation}
\frac{1}{n!}\sum_{\sigma\in S_n} 1 \chi^{\lambda}(\sigma) = \sum_{\sigma\in S_n} \chi^{(n)}(\sigma)\chi^{\lambda}(\sigma) = \delta_{(n),\lambda} \, .
\end{equation}
Substituting into \eref{Jnunitproofbegin} the gives
\begin{equation} 
\J_n = \frac{1}{n!} \sum_{\lambda \vdash n} \frac{f^{\lambda}}{C_\lambda(N)} \delta_{(n),\lambda} = \frac{f^{(n)}}{C_{(n)}(N)} \, .
\end{equation}
Since $f^{(n)}=1$ and $C_{(n)}(N)=N(N+1)\ldots(N+n-1)$ from the definitions in 
\cite{matsumoto13} we obtain
\begin{equation}
\J_n = \frac{1}{N(N+1)\ldots(N+n-1)} \, ,
\end{equation}
recovering and proving \eref{Jnallnunit}.

\subsubsectionSupp{Unitary fermions}

For fermions we need to include the powers of $(-1)$ in \eref{Bsquaresimp}.  For this we proceed as in the bosonic case, but now we for the powers of $(-1)$ we use the irreducible character $\chi^{(1^n)}(\sigma)=(-1)^{\sigma}$ for $\sigma\in S_n$.  Substituting into \eref{Bsquaresimp} gives
\begin{equation} 
\Jf_n = \frac{1}{n!} \sum_{\lambda \vdash n} \frac{f^{\lambda}}{C_\lambda(N)} \sum_{\sigma\in S_n} \chi^{(1^n)}(\sigma) \chi^{\lambda}(\sigma) \, ,
\end{equation}
while orthogonality reduces the result to
\begin{equation} 
\Jf_n = \frac{1}{n!} \sum_{\lambda \vdash n} \frac{f^{\lambda}}{C_\lambda(N)} \delta_{(1^n),\lambda} = \frac{f^{(1^n)}}{C_{(1^n)}(N)} \, .
\end{equation}
Taking $f^{(1^n)}=1$ and $C_{(1^n)}(N)=N(N-1)\ldots(N-n+1)$ from the definitions in 
\cite{matsumoto13} we obtain
\begin{equation}
\Jf_n = \frac{1}{N(N-1)\ldots(N-n+1)} \, ,
\end{equation}
proving \eref{Jfnallnunit}.

\subsubsectionSupp{Orthogonal fermions}

This proof is somewhat more involved and we start by expressing our sum
\begin{eqnarray}
\Jf_n &=& \sum_{\tau \in S_n} (-1)^{\tau} V_{N}(\tau) \nonumber \\
&=& \sum_{\mu \vdash n} \frac{n!}{z_\mu} (-1)^{n-l(\mu)} \Wg(\mu; N+1) \, ,
\end{eqnarray}
in terms of $\Wg$ which are the Weingarten function for the COE and which are evaluated for permutations $\tau$ of coset-type $\mu$ while $z_\mu$ is as defined in \cite{matsumoto12}.  As in \cite{matsumoto12} we can reexpress our sum in terms of double length permutations 
\begin{equation}
\Jf_n = \frac{1}{2^n n!}\sum_{\sigma \in S_{2n}}\left(-\frac{1}{2}\right)^{n-l'(\sigma)} \Wg(\sigma; N+1)\, ,
\end{equation}
where $l'(\sigma)$ is the length of $\mu$ if $\mu$ is the coset-type of $\sigma$.  From the definition of the orthogonal Weingarten function \cite{matsumoto13} our sum becomes
\begin{equation}
\Jf_n = \frac{1}{(2n)!}\sum_{\lambda \vdash n}\frac{f^{2\lambda}}{D_{\lambda}(N+1)}\sum_{\sigma \in S_{2n}}\left(-\frac{1}{2}\right)^{n-l'(\sigma)} \omega^{\lambda}(\sigma)\, ,
\end{equation}
in terms of zonal spherical functions $\omega^{\lambda}$.  These play the role of the irreducible characters used for the unitary case, and analogously as for the unitary fermions
\begin{equation}
\omega^{(1^n)}(\sigma) = \left(-\frac{1}{2}\right)^{n-l'(\sigma)} \, .
\end{equation}
The zonal functions also follow an othogonality relation
\begin{equation}
\frac{1}{(2n)!}\sum_{\sigma\in S_{2n}} \omega^{\lambda}(\sigma)\omega^{\mu}(\sigma) = \frac{\delta_{\lambda,\mu}}{f^{2\lambda}} \, ,
\end{equation}
so that
\begin{equation}
\Jf_n = \sum_{\lambda \vdash n}\frac{f^{2\lambda}}{D_{\lambda}(N+1)}\frac{\delta_{\lambda,(1^n)}}{f^{2\lambda}} = \frac{1}{D_{(1^n)}(N+1)} \, .
\end{equation}
Finally from the definition of $D_\lambda(N+1)$ in \cite{matsumoto13} one has $D_{(1^n)}(N+1)=\prod_{i=1}^{n}(N+1-i)$ giving
\begin{equation}
\Jf_n = \frac{1}{(N+1)N\ldots(N-n+2)} \, ,
\end{equation}
proving \eref{Jfnallnorth}.

\subsectionSupp{Coinciding channels}

We start with letting $k$ outgoing channels (say $b_i$, $i=1,\ldots,k$) be identical while keeping the remaining outgoing channels and the incoming channels distinct.  Then still only $\sigma=\id$ is permissible in \eref{scatcorrel} while $\pi$ can now take any value $\K\P$ for all $\K\in S_k$.  The sum becomes
\begin{equation} \label{Asquarecoinc}
\Jhbf_n = \sum_{\P\in S_n}(\pm1)^{\P} \sum_{\K\in S_k} V_{N}(\K\P) \, .
\end{equation}
For each $\K$ we set $\tau=\K\P$ then since
\begin{equation} 
(\pm1)^{\P} = (\pm1)^{\K^{-1}\tau} = (\pm1)^{\K^{-1}}(\pm1)^{\tau} = (\pm1)^{\K}(\pm1)^{\tau} \, ,
\end{equation}
the sum reduces to
\begin{equation}
\Jhbf_n = \sum_{\K\in S_k} (\pm1)^{\K}\sum_{\tau\in S_n}(\pm1)^{\tau}  V_{N}(\tau) \, .
\end{equation}
Since we already know the sum over $\tau$
\begin{equation}
\Jhbf_n = \sum_{\K\in S_k} (\pm1)^{\K} \Jbf_n \, ,
\end{equation}
we are left with the simple sum over $\K$.  For bosons, this is simply $k!$ while for fermions we can again use the irreducible characters and their orthogonality
\begin{equation}
\sum_{\K\in S_k} (-1)^{\K} = \sum_{\K\in S_k} \chi^{(1^k)}(\K) \chi^{(k)}(\K) = k! \delta^{(1^k),(k)} = \delta_{k,1} \, ,
\end{equation}
since clearly $(1^k)$ and $(k)$ can only be the same partition when $k=1$.  Combined we have 
\begin{equation}
\Jhbf_n = k! (1\pm1) \Jbf_n \, \qquad k>1  \, .
\end{equation}

We can repeat this process for arbitrary sets of coinciding incoming and outgoing channels giving the result for bosons that $\Jh_n = {\bf a}!{\bf b}! \J_n$ and zero for fermions as soon as any channels coincide.  For the unitary case there is no restriction on whether ${\bf a}$ or ${\bf b}$ contain the same channels, but in the orthogonal case as soon as this happens the simple formula in \eref{scatcorrel} is no longer valid and must be replaced by a more complicated version (see \cite{bb96,bk13a} for example).  With this restriction, these results provide \eref{eq:Pconj} in the main text.

\sectionSupp{Semiclassical treatment of scattering matrix correlators}
\label{sec:SuppJI}

We will treat correlators of $A_n$ using a semiclassical diagrammatic approach.  This is heavily based on \cite{bk11,bk12,bk13a,bk13b} and we refer in particular to \cite{bk11,bk13a} for the underlying details and methods.

\begin{figure*}
  \centering
  \includegraphics[width=\textwidth]{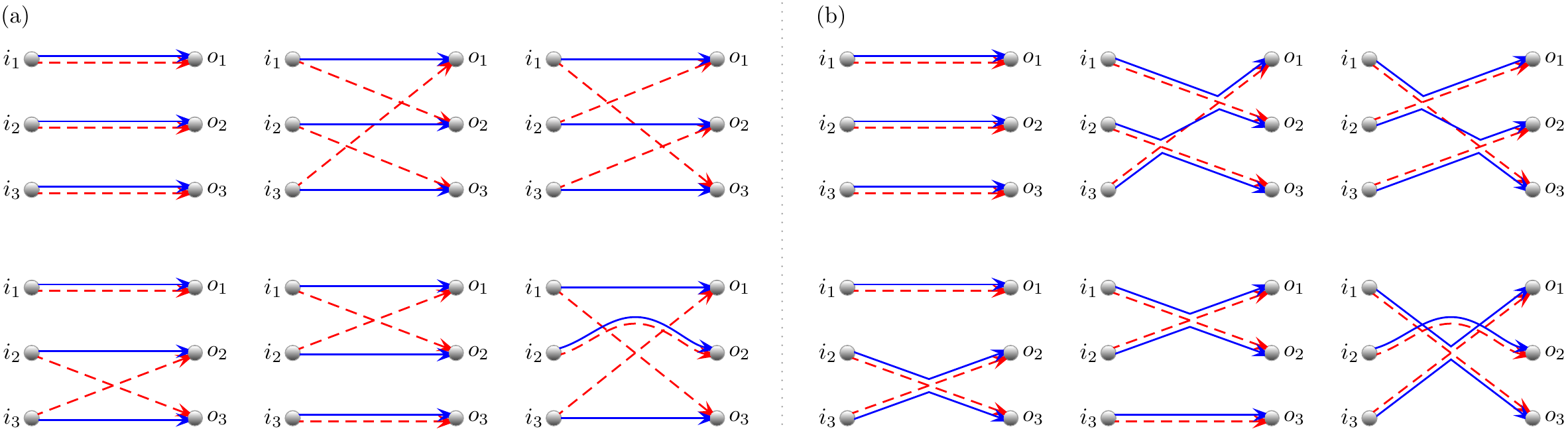}
  \caption{(a) The permutations on 3 labels represented as trajectory diagrams. (b) Semiclassical contributions come when the trajectories are nearly identical, as when collapsed onto each other.}
  \label{fig:nis3diags}
\end{figure*}

We return first to the transmission probability for bosons in \eref{Asquaresimpler}.  For a given cycle $(1,\ldots,l)$ in the target permutation $\tau$ the semiclassical trajectories have a very particular structure whereby we first travel along a trajectory with positive action from $i_1$ to $o_1$ and then in reverse back along a trajectory with negative action to $i_2$ and so on along a cycle until we return to $i_1$.   For example for $n=3$ we have the trajectory connections in \fref{fig:nis3diags}(a) for each target permutation $\tau$ in \eref{nis3perms}.

For the actions of the diagram to nearly cancel, and to obtain a semiclassical contribution, the trajectories must be nearly identical, except at small regions called encounters.  By directly collapsing the trajectories onto each other, as in \fref{fig:nis3diags}(b) we obtain some of the leading order diagrams for each $\tau$.  In fact for each diagram, following the rules of \cite{mulleretal07}, the semiclassical contribution is a factor of $-N$ for each encounter and a factor of $N^{-1}$ for each link between the encounters. For each cycle of length $l$ in the diagrams in \fref{fig:nis3diags}(b) one then has a factor of order $N^{-2l+1}$.

As a straightforward example we can look at the simplest diagrams made up of a set of independent links like the first diagram in \fref{fig:nis3diags}(b).  With $n$ links each providing the factor $N^{-1}$ we have the contribution 
\begin{equation}
\label{eq:JsuppI}
\Jbf = N^{-n}
\end{equation}
This contribution is in fact unaffected by the energy and time differences of the incoming particles leading directly to the classical contribution presented in \eref{eq:PclassFA} of the main text.  This contribution also accounts for the leading order terms in the expansions of Eqs.~(\ref{Jnunitexpand}),  (\ref{Jnorthexpand}), (\ref{Jfnunitexpand}) and (\ref{Jfnorthexpand}).

\subsectionSupp{Diagrammatic treatment without time reversal symmetry}

Once the contribution of each diagram has been established, one then needs to generate all permissible diagrams.  As shown in \cite{bk12,bk13a} however the vast majority of semiclassical transport diagrams cancel.  Those which remain can be untied until their target permutation becomes identity.  For systems without time reversal symmetry, which we consider first, they can be mapped to primitive factorisations.  One can reverse the process to build the diagrams by starting with a set of $n$ independent links and tying together two outgoing channels into a new encounter.  This tying process increases the order of the diagram by $N^{-1}$.  If the outgoing channels are labelled by $j$ and $k$ then the target permutation also changes to $\tau (j\,k)$.  For example going from the top left diagram of \fref{fig:nis3diags}(b) tying together any two outgoing channels leads to the three example along the bottom row.  The diagrams correspondingly move from order $N^{-3}$ to $N^{-4}$.

Tying the remaining outgoing channel to one of those already tied leads to a diagram of the type further along the top row of \fref{fig:nis3diags}(b) [for each of which there are 3 possible arrangements, and an alternative with a single larger encounter] and now of order $N^{-5}$. 

Of course one could retie the same pair chosen in the first step, so that the target permutation is again identity.  Such a diagram is however not shown in \fref{fig:nis3diags}(b) but can be thought of as a higher order correction to a diagonal pair of trajectories.  These types of diagrams appear when one treats the conductance variance for example.  Such diagrams have a graphical interpretation which we will discuss below and use to generate them.

\subsubsectionSupp{Forests}

At leading order for each cycle of length $l$ in $\tau$ the trajectories however form a ribbon graph in the shape of a tree.  The tree has $2l$ leaves (vertices of degree 1) and all further vertices of even degree greater than 2.  Such trees can be generated \cite{bhn08} by first treating unrooted trees whose contributions we store in the generating function $f$.  Using the notation in \cite{bk11}, the function satisfies
\begin{equation} \label{feqn}
f = \frac{r}{N} - \sum_{k=2}^{\infty} f^{2k-1} \, , \qquad \frac{f}{N}=\frac{\sqrt{1+\frac{4r^2}{N^2}}-1}{2r} \, ,
\end{equation}
where the power of $r$ counts the number of leaves and the encounters may not touch the leads since the channels are distinct.  Rooting the tree we add a leave to arrive at the generating function $F=rf$ while setting $r^2=s$ we arrive at
\begin{equation}
\frac{F}{N} = \frac{\sqrt{1+\frac{4s}{N^2}}-1}{2} \, .
\end{equation}
Expanding in powers of $s$ 
\begin{equation}
F = \frac{s}{N} - \frac{s^2}{N^3} + \frac{2s^3}{N^5} - \frac{5s^4}{N^7} + \frac{14s^5}{N^9} + \ldots
\end{equation}
one has an alternating sequence of Catalan numbers, A000108 \cite{sloanes}.

When summing over all permutations for $\tau$, each cycle of length $l$ can be arranged in $(l-1)!$ ways and we now wish to include this factor in the ordinary generating function.  First we divide instead by a factor $l$ with the transformation
\begin{eqnarray}
\label{eq:JGenF}
\frac{\hF}{N} = \int \frac{F}{sN} \, \rmd s &=&  \sqrt{1+\frac{4s}{N^2}} - 1 \\ 
& & {} +
\frac{1}{2}\ln\left[\frac{N^4\left(1-\sqrt{1+\frac{4s}{N^2}}\right)}{2s^2}+\frac{N^2}{s}\right] \, , \nonumber
\end{eqnarray}
so that $\hF$ becomes the exponential generating function of the leading order trees multiplied by the factor $(l-1)!$ as required.  To now generate any forest of trees corresponding to all permutations $\tau$ we can exponentiate $\hF$ to obtain the exponential generating function
\begin{eqnarray} \label{Geqn}
\rme^{\hF} - 1  & = &\frac{s}{N}+\frac{(N-1)s^2}{2N^3}+\frac{(N^2-3N+4)s^3}{6N^5} \nonumber \\
& & {} + \frac{(N^3-6N^2+19N-30)s^4}{24N^7} + \ldots
\end{eqnarray}
whose first few terms can be explicitly checked against diagrams.

\subsubsectionSupp{Higher order corrections to trees}

For each given cycle $(1,\ldots,l)$ of $\tau$ there are higher order (in $N^{-1}$) corrections which can be organised in a diagrammatic expansion \cite{bk11,bk13b}.  For systems without time reversal symmetry, the first correction occurs two orders lower than leading order and the corresponding diagrams can be generated by grafting the unrooted trees on two particular base diagrams.  Repeating the steps in \cite{bk11}, while excluding the possibility for encounters to touch the leads (since the channels are distinct), one first obtains
\begin{equation}
\Ktwo=-\frac{(f^2+3)f^4}{6(f^2+1)^3} \, ,
\end{equation}
where handily, the method for subleading corrections automatically undercounts by a factor of $l$ so we directly obtain the required exponential generating function.  Finally we substitute from \eref{feqn} and find
\begin{equation}
12N\Ktwo =  \frac{1+\frac{6s}{N^2}}{\left(1+\frac{4s}{N^2}\right)^{\frac{3}{2}}}- 1 \, .
\end{equation}
The exponential generating function $\rme^{\hF+\Ktwo} - 1$ would then generate all corresponding diagram sets up to this order.

\subsubsectionSupp{Other corrections}

However, the higher order corrections to trees are less important than the higher order corrections to other target permutation structures.  For any pair of cycles $(1,\ldots,k)(k+1,\ldots,l)$ in $\tau$ we can have diagrams which are order $N^{-2}$ smaller than a pair of leading order trees.  For example, tying any two outgoing channels of a tree on the cycle $(1,\ldots,l)$ would break the target permutation into two as here.  

To generate diagrams with two cycles, we graft trees around both sides of a circle as for the cross correlation of transport moments treated in \cite{bk11}.  This will include the example with $n=3$ mentioned at the start of this subsection.   
 
Following the steps in \cite{bk11}, while excluding the possibility of encounters touching the lead, one finds the generating function
\begin{eqnarray}
\kappa &=& -\ln\left[\frac{1-f_1^2f_2^2}{(1-f_1^2)^2(1-f_2^2)^2}\right] \\ 
& & {} + \ln\left[\frac{1}{(1-f_1^2)^2}\right] + \ln\left[\frac{1}{(1-f_2^2)^2}\right] \, , \nonumber
\end{eqnarray}
where $f_1$ and $f_2$ are the $f$ in \eref{feqn} but with arguments $r_1$ and $r_2$ respectively.  The last two terms are corrections for when either $r_1$ or $r_2$ is 0 to remove diagrams with no trees on either side of the circle.  In \cite{bk11}, $\kappa$ was differentiated and such terms removed automatically, but here this correction simplifies the result to
\begin{equation}
\kappa = -\ln\left[1-f_1^2f_2^2\right] \, .
\end{equation}
This generating function again undercounts by both a factor of $k$ and $(l-k)$ and can therefore be thought of as an exponential generating function of both arguments.  Setting $s_1=s_2=s$ then sums the possible splittings of the $l$ elements into two cycles (along with the combinatorial factor of choosing the label sets), counting each splitting twice.  This then provides the following exponential generating function
\begin{eqnarray}
\kk &=& -\frac{1}{2}\ln\left[1-f^4\right] \\ 
&=& -\frac{1}{2}\ln\left[\frac{N^4}{2s^2} \left(\sqrt{1+\frac{4s}{N^2}}\left(1+\frac{2s}{N^2}\right)-1-\frac{4s}{N^2}\right)\right] \, , \nonumber
\end{eqnarray}
whose expansion is
\begin{equation}
\kk = \frac{s^2}{2N^4}-\frac{2s^3}{N^6}+\frac{29s^4}{4N^8}-\frac{26s^5}{N^{10}}+\ldots
\end{equation}

\subsubsectionSupp{Order of contributions}
\label{sec:SuppJIIIA4}

With the diagrams treated so far the exponential generating function would be
\begin{eqnarray}
\rme^{\hF+\kk+\Ktwo} - 1 & = & \frac{s}{N}+\frac{(N^3-N^2+N-1)s^2}{2N^5} \nonumber \\ 
& & {} +\frac{(N^4-3N^3+7N^2-15N+20)s^3}{6N^7} \nonumber \\
& & {} +\ldots
\end{eqnarray}
and the differences from \eref{Geqn} occur two orders lower in $N^{-1}$ than the leading term from independent links.  This is because the additional diagrams required at least two tying operations.  However, we wish to know how the contributions change when $n$ scales with $N$ in some way.

First we can compare the contributions coming from $\Ktwo$ to those from $\hF$.  In the forest we can replace any tree from $\hF$ with its higher order correction in $\Ktwo$.  Since there can be at most $n$ trees, and the correction is order $N^{-2}$ smaller, these corrections will be bound by $nN^{-2}$, up to the scale of the generating function coefficients.  This means that we can expect the contributions from $\Ktwo$ to not be important when $n=o(N^2)$ as we take the limit $N\to\infty$.

Next we compare the contributions coming from $\kk$ to those from $\hF$.  In the forest we can now replace any tree by breaking its $l$ cycle into two, say $k$ and $(l-k)$.  Alongside the generating function coefficients, the two new cycles come with the factor $(l-k-1)!(k-1)!$ instead of the $(l-1)!$ that was with the tree. Since
\begin{eqnarray}
& & \frac{1}{(l-1)!}\sum_{k=1}{l-1}\left(\begin{array}{c} l \\ k \end{array}\right) (l-k-1)!(k-1)! \nonumber \\
&=& \sum_{k=1}^{l-1}\frac{l}{(l-k)k} \leq l \, ,
\end{eqnarray}
this contribution is bound by $nN^{-1}$ and should not be important when $n=o(N)$.

Continuing in this vein we could break up any tree into three cycles and generate those diagrams but this should also be higher order when $n\ll N$. Keeping our expansion to this order we should have
\begin{eqnarray}
\rme^{\hF+\kk} - 1 & = & \frac{s}{N}+\frac{(N^2-N+1)s^2}{2N^4} \\ 
& & {} +\frac{(N^3-3N^2+7N-12)s^3}{6N^6} +\ldots \nonumber 
\end{eqnarray}

\subsectionSupp{Time reversal symmetry}

With time reversal symmetry, additional diagrams become possible.  For example we may reverse the trajectories on one side of the circle used for cross correlations and obtain $2\kk$ instead of just $\kk$.  There are also additional base diagrams at the second order correction to trees, which may be treated as in \cite{bk11}, but which we do not treat here since there are now diagrams at the the first order correction.  These can be generated by grafting trees around a M\"obius strip.  Following again the steps in \cite{bk11} while excluding the possibility of encounters touching the lead one obtains the generating function
\begin{equation}
\Kone = \frac{1}{2}\ln\left[\frac{1-f^2}{1+f^2}\right] \, ,
\end{equation}
or explicitly
\begin{eqnarray}
\Kone &=& -\frac{1}{4}\ln\left[1+\frac{4s}{N^2}\right] \\
&=& -\frac{s}{N^2} + \frac{2s^2}{N^4} - \frac{16s^2}{3N^6} + \frac{16s^4}{N^8} - \frac{256s^5}{5N^{10}} + \ldots \nonumber
\end{eqnarray}

Compared to the leading order forest, we could replace any tree by its higher order correction and obtain a term bound by $nN^{-1}$, again up to the scale of the generating function coefficients.  Restricting to $n=o(N)$ the the exponential generating function would be
\begin{eqnarray}
\rme^{\hF+2\kk+\Kone} - 1 & = & \frac{(N-1)s}{N^2}+\frac{(N^2-3N+7)s^2}{2N^4} \nonumber \\ 
& & {} +\frac{(N^3-6N^2+28N-75)s^3}{6N^6} \nonumber \\
& & {} +\ldots
\end{eqnarray}

\subsectionSupp{Fermions}

For fermions we need to also include the powers of $(-1)$ in \eref{Bsquaresimp}.  However, because our semiclassical generating functions are organised by cycle type, we simply need to replace $s$ by $-s$ and multiply the $K$ type functions by -1 appropriately.

\subsectionSupp{The variance}
\label{sec:SuppVar}

\begin{figure*}
  \centering
  \includegraphics[width=\textwidth]{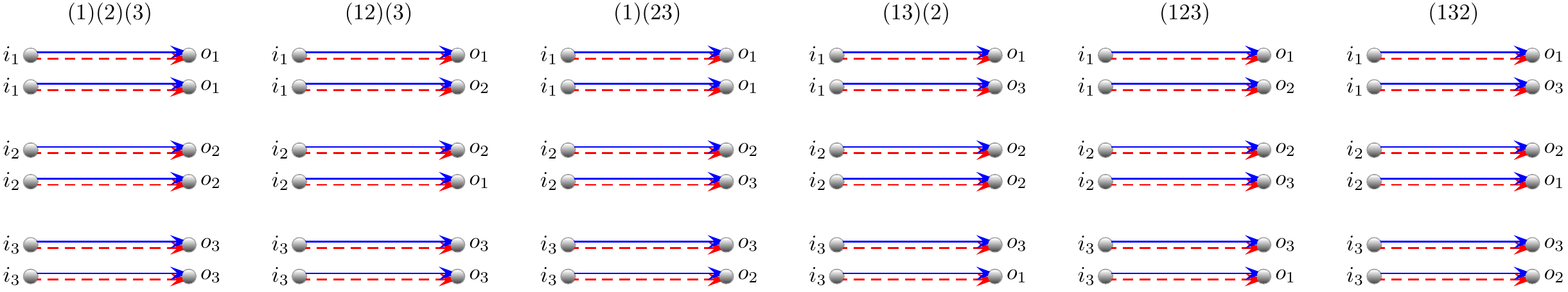}
  \caption{The leading order diagrams for the variance with 3 particles are created by finding semiclassical diagrams with 6 incoming and outgoing channels.  When the channels coincide, the leading order diagrams must reduce to separated links corresponding to one of the permutations on 3 labels depicted.}
  \label{fig:nis3varlo}
\end{figure*}

Now for $n$ bosons we look at
\begin{eqnarray} \label{Aquaddef}
\vert \A^{+}_n\vert^4 &=& \frac{1}{(n!)^2} \sum_{\substack{\P,\P' \cr \R,\R'}\in S_n} \prod_{k=1}^n Z_{i_k,o_{\P(k)}}Z^{*}_{i_{k},o_{\P'(k)}} \nonumber \\
& & \times Z_{i_k,o_{\R(k)}}Z^{*}_{i_{k},o_{\R'(k)}} \, ,
\end{eqnarray}
or rather the average
\begin{equation} \label{Ldef}
\L_n = \left\langle \vert \A^{+}_n\vert^4 \right\rangle \, .
\end{equation}
However, when we now compare to \eref{scatcorrel} such an average involves summing over permutations of length $2n$ while each of the originally distinct channels appears exactly twice.  For example
\begin{equation} \label{L1}
\L_1 = \langle Z_{i_1,o_1}Z_{i_1,o_1}Z^{*}_{i_1,o_1}Z^{*}_{i_1,o_1} \rangle=2V_{N}(1,1)+2V_{N}(2) \, ,
\end{equation}
since the delta function conditions in \eref{scatcorrel} are satisfied for any choice of $\sigma$ and $\pi$.  The result is 
\begin{equation} \label{L1result}
\L_1 = \frac{2}{N(N+1)} \, \qquad \L_1=\frac{2}{N(N+3)} \, ,
\end{equation}
without or with time reversal symmetry respectively.

For $n=2$, we can run through the sums of permutations, giving 
\begin{equation} \label{L2resultunit}
\L_2 = \frac{3N^2-N+2}{N^2(N^2-1)(N+2)(N+3)} \, ,
\end{equation}
without time reversal symmetry and
\begin{equation} \label{L2resultorth}
\L_2=\frac{3N^2+5N-16}{N(N^2-4)(N+1)(N+3)(N+7)} \, ,
\end{equation}
with. For large $n$ this process however quickly becomes computationally intractable.  Diagrammatically, we can imagine multiplying the sets of diagrams we had before, but also keeping track of all the possible permutations and which channels coincide.  For example we could take the diagrams in \fref{fig:nis3diags}(b), add the remaining 5 copies of each diagram created by permuting the outgoing labels, and multiply the entire set by itself to obtain pairs of diagrams which appear for the variance.  Of course each pair is over counted $(n!)^2$ times and we would still need to account for the diagrams where the pairs interact and where the repeated channels play a role by considering diagrams acting on $2n$ leaves.

To reduce the difficulty of such a diagrammatic expansion, we focus here instead on just calculating the leading order term.  We know that these terms are represented diagrammatically by sets of independent links so we select the $(n!)^2$ such sets from our multiplication.  Since each outgoing channel (although appearing twice) is distinct we may relabel them appropriately to reduce our leading order diagrams to $n!$ ways of permuting a single outgoing label. The sum of a product of two permutations essentially reduces to a sum over a single permutation.  The overcounting is now $n!$ instead. For $n=3$ the leading order diagrams are depicted in \fref{fig:nis3varlo}.  For each diagram we have the standard leading order result of $N^{-2n}$ which, when dividing by the over counting, would be the total result if the outgoing channels were different.

\begin{figure*}
  \centering
  \includegraphics[width=\textwidth]{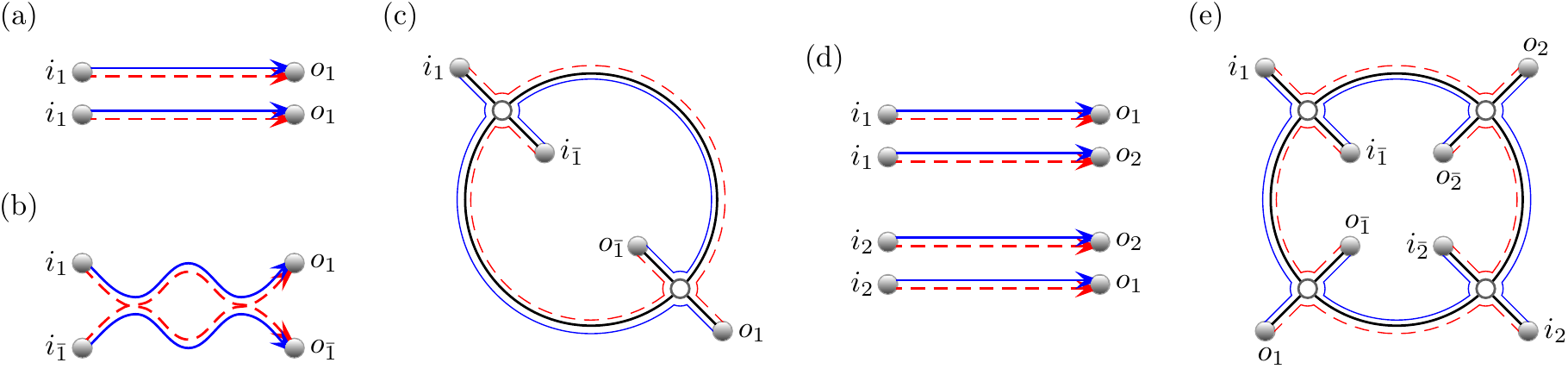}
  \caption{(a) A pair of independent links can be joined by an encounter at each end to create the diagram in (b) with artificially distinct incoming and outgoing channels.  When the encounters are moved into the incoming and outgoing leads respectively, the channels again coincide leading to a new leading order semiclassical diagram.  In the graphical representation, the trajectories in (b) become the boundary walks around both sides of the circle in (c).  Starting with four links corresponding to the permutation $(12)$ in (d) we can create the correlated quadruplet represented in (e).  Again moving the encounters into the leads creates a new leading order contribution.}
  \label{fig:locycles}
\end{figure*}

However, for each cycle of the effective permutation of the outgoing channels, an additional semiclassical diagram is possible.  By adding a 2-encounter to each pair of identical channels we can separate them into two artificially distinct channels.  The resulting semiclassical diagram can be drawn as a series of 2-encounters around a circle with one link on either side.  This process is depicted in \fref{fig:locycles}.  The resulting starting point is from the larger set of possible trajectory correlators than just the sets of independent links squared, but once we move all the encounters into the appropriate leads, the required channels coincide and we have an additional leading order diagram.

To count such possibilities we just need to include a power of 2 for each cycle in the permutation of the outgoing channels when we include the standard diagonal terms.  For each cycle of length $l$ there are $(l-1)!$ different permutations so that 
\begin{equation} 
-2\log(1-s) = 2s+s^2 + \frac{2s^3}{3}+\frac{s^4}{2}+\ldots
\end{equation}
acts as the exponential generating function of both possibilities for each cycle times their number of permutations.   To generate all leading order diagrams we simply exponentiate this function
\begin{equation} 
\rme^{-2\log(1-s)}-1 = \frac{1}{(1-s)^2}-1
\end{equation}
Since we are still overcounting by $n!$ this actually provides the ordinary generating function and when we include the semiclassical contributions of $N^{-2n}$ we get the final leading order result of 
\begin{equation} \label{Lnleadingorder}
\L_n = \frac{n+1}{N^{2n}} + o(N^{-2n+1})
\end{equation}
or a variance of
\begin{equation} \label{varleadingorder}
\L_n -\left(\J_n\right)^2 = \frac{n}{N^{2n}} + o(N^{-2n+1})
\end{equation}
We checked this against the explicit semiclassical or RMT results involving \eref{scatcorrel} for $n$ up to 5.  Since the semiclassical diagrams all involve pairs of equally long cycles, the leading order result is also the same for fermions.

Intriguingly, the numerator of the leading order result for the second moments in \eref{Lnleadingorder} is identical to the second moment of the modulus squared of the permanents of $n\times n$ random complex Gaussian matrices \cite{aa13}.  Higher moments of such permanents would be useful to determine the validity of the {\it Permanent Anti-Concentration Conjecture} important for Boson Sampling \cite{aa13}.  This opens the possibility that a semiclassical or RMT treatment of the higher moments of many body scattering, expanded just to leading order, could help answer such questions.

\sectionSupp{Relevance of diagrams with pairwise correlations in the scaling limit}
\label{sec:SuppPair}
In the following sections we rigurously shown that in the limit $n\to \infty$ and under the scaling $N=\alpha n^{2}$ the sum of semiclassical diagrams with pairwise correlations gives Eq.~(\ref{eq:finII}) of the main text if $\tau_{d}/\tau_{s} = 0, z_{ij} = 0$. Once this result is stablished, we re-introduce $\tau_{d}/\tau_{s} > 0, z_{ij} \ne 0$ and obtain Eq.~(\ref{eq:finI}), one of our main results.

\subsectionSupp{The case $\tau_{d}/\tau_{s} = 0, z_{ij} = 0$}
\label{sec:SuppJJDI}
In order to show that only diagrams with pairwise correlations are neccesary to obtain Eq.~(\ref{eq:finII}) in the main text, we start with the following exact relation
\begin{equation}
\label{eq:SuppJDJGen}
\left.\frac{\langle P^{(\epsilon)}_{{\bf a},{\bf \underline{b}}}\rangle}{\langle P^{{\rm (cl)}}_{{\bf a},{\bf \underline{b}}}\rangle}\right|_{z_{ij}=0}^{\frac{\tau_{d}}{\tau_{s}}=0}=\frac{\partial^{n}}{\partial s^{n}}\left.{\rm e}^{K_{0}(s)}\right|_{s=0}
\end{equation}
expressing the ratio between quantum and classical transition probabilities in terms of the generating function $K_{0}(s)$ defined in Eq.~(\ref{eq:JGenF}). In the limit $n,N \to \infty$ with quadratic scaling $N=\alpha n^{2}$, all bounds in Sec.~\ref{sec:SuppJIIIA4} show we only need to consider standard trees and forests. The generating function $K_0$ generates trees, with each power of $s$ corresponding to trees with increasing number of leaves.  Just $s$ by itself is individual links, $s^2$ are the pairwise correlations, $s^3$ would be all three way correlator and so on. If we truncate to second order we only have links and x-like correlations in our generating function while further exponentiating $K_0$ truncated to second order generates all possible sets of links and pairwise correlators, like (a), (b), and (e) but not (d) in Fig.~\ref{fig:jack} of the main text.

Our strategy is to show that, in the scaling limit, Eq.~(\ref{eq:SuppJDJGen}) gives Eq.~(\ref{eq:finII}) of the main text when $K_{0}=s-s^{2}/2N+{\cal O}(s^{3})$ is truncated to second order and therefore only pairwise correlations are included. Our starting point is then Eq.~(\ref{eq:SuppJDJGen}) with  $K_{0}=s(1-s/2N)$,
\begin{equation}
\left.\frac{\langle P^{(\epsilon)}_{{\bf a},{\bf \underline{b}}}\rangle}{\langle P^{{\rm (cl)}}_{{\bf a},{\bf \underline{b}}}\rangle}\right|_{z_{ij}=0}^{\frac{\tau_{d}}{\tau_{s}}=0} \simeq \frac{\partial^{n}}{\partial s^{n}}\left.{\rm e}^{s\left(1-\frac{s}{2N}\right)}\right|_{s=0},
\end{equation}
which can be written in the convenient form for asymptotic analysis
\begin{eqnarray} 
\label{eq:saddle}
\frac{\partial^{n}}{\partial s^{n}}\left.{\rm e}^{s\left(1-\frac{s}{2N}\right)}\right|_{s=0}&=&\frac{n!}{2\pi i}\oint \frac{{\rm e}^{z\left(1-\frac{z}{2N}\right)}}{z^{n+1}} \\ &=&\frac{n!}{2\pi i}\oint {\rm e}^{z\left(1-\frac{z}{2N}\right)-(n+1)\log z} \nonumber
\end{eqnarray}
as a complex integral along a contour that enlcoses the origin. In the large $n$ limit this integral can be evaluated in saddle point approximation. The saddle point $z=z^{c}$ is easily found to be
\begin{equation}
z^{c}=\frac{N}{2}\left(1-\sqrt{1-\frac{4(n+1)}{N}}\right) \simeq n+1+\frac{(n+1)^{2}}{N}.
\end{equation}
The exponent in Eq.~(\ref{eq:saddle}) and its second derivative evaluated at the saddle point are also easily found to be
\begin{eqnarray}
&&\left.\left(z\left(1-\frac{z}{2N}\right)-(n+1)\log z\right)\right|_{z=z^{c}}\simeq -(n+1)\log(n+1)\nonumber \\ && +(n+1)-\frac{(n+1)^{2}}{2N}, \\
&&\left. \frac{\partial^{2}}{\partial z^{2}}\left(z\left(1-\frac{z}{2N}\right)-(n+1)\log z \right)\right|_{z=z^{c}}\simeq\frac{1}{n+1}\nonumber
\end{eqnarray}
to get
\begin{equation}
\frac{\partial^{n}}{\partial s^{n}}\left.{\rm e}^{s\left(1-\frac{s}{2N}\right)}\right|_{s=0}\simeq \frac{n!}{2\pi}\frac{\sqrt{2\pi (n+1)}}{((n+1)/{\rm e})^{n+1}}{\rm e}^{-\frac{(n+1)^{2}}{2N}}
\end{equation}
and finally, using the asymptotic approximation for the factorial of large numbers,
\begin{equation}
\frac{\partial^{n}}{\partial s^{n}}\left.{\rm e}^{s\left(1-\frac{s}{2N}\right)}\right|_{s=0}\simeq {\rm e}^{-\frac{n^{2}}{2N}}.
\end{equation}

We conclude then that in the limit $n\to \infty$ and under the scaling $N=\alpha n^{2}$ the sum of semiclassical diagrams with pairwise correlations gives Eq.~(\ref{eq:finII}) of the main text.

\pagebreak

\subsectionSupp{The case $\tau_{d}/\tau_{s} > 0, z_{ij} \ne 0$}
\label{sec:SuppJJDII}

If only pairwise correlations are included in the diagramatic expansion, all results are expressed in terms of the overlapping functions between all posible pairs of incoming channels. This is a simple combinatorial problem and we get (for $\eta > 1$)
\begin{equation}
\label{eq:P22}
\frac{\langle P^{(\epsilon)}_{{\bf a},{\bf b}}\rangle}{\langle P^{{\rm (cl)}}_{{\bf a},{\bf b}}\rangle}\xrightarrow[{N=\alpha n^{\eta}}]{n\gg1}
	\sum_{\substack{\mathfrak{I} \subseteq \{1,\ldots,n\} \\ \lvert\mathfrak{I}\rvert =: 2l \ {\rm even}} }
	\frac{ (-\epsilon)^l }{N^l} \sum_{{\cal C} \sqcap \mathfrak{I}}\prod_{q=1}^{l}{\cal Q}^{(2)}(z_{{\cal C}_{q}}) \,,
\end{equation}  
where ${\cal C}$ runs over the set of contractions obtained by pairing different indexes in $\mathfrak{I}=\{i_{1},\ldots, i_{2l}\}$ and ${\cal C}_{q}$ is its $q$th element.

\subsectionSupp{Mesoscopic HOM effect with two macroscopically occupied channels}
\label{sec:SuppH}
In the following the explicit evaluation of Eq.~(\ref{eq:P22}) in the case of a very large number $n \gg 1$ of particles that macroscopically occupy only two different incoming wavepackets is carried out.
Let $n_0 = y_0 n$ denote the number of particles in a single incoming wavepacket for which we set $z=0$ and let $n_1 = y_1 n$ be the number of remaining particles occupying a wavepacket with separation $z=z_1$ from the other.
We denote the corresponding sets of indexes with $\mathfrak{I}_0$ and $\mathfrak{I}_1$, respectively.

Pairings of the $n$ particle indices are characterized by the total number of pairs $l$.
Furthermore this number splits into the number $l_1$ of pairs among the $n_0$ indexes in $\mathfrak{I}_0$, the number $k$ of pairs connecting an index in $\mathfrak{I}_0$ with one in $\mathfrak{I}_1$ and the number $l-l_1-k$ of pairs inside $\mathfrak{I}_1$.
Each contraction specified by the numbers $l,l_1,k$ contributes a value of
\begin{equation}
\label{eq:SuppQin}
	\left( \frac{-\epsilon}{N} \right)^l [ {\cal Q}^{(2)}(0) ]^{l-k} [ {\cal Q}^{(2)}(z_1) ]^k =: \left( \frac{-\epsilon}{\alpha} \right)^l n^{-2l} q_0^{l-k} q_1^{k}
\end{equation}
to the sum in Eq.~(\ref{eq:P22}) if the scaling $N=\alpha n^2$ is taken into account.
We introduced the abbreviations $q_0$ and $q_1$ for the two overlap integrals involved.
Therefore the probability to get all particles in different outgoing channels is
\begin{equation} \label{eq:Pinfty1}
	\begin{split}
		\frac{\langle P^{(\epsilon)}_{{\bf a},\underline{{\bf b}}}\rangle}{\langle P^{{\rm (cl)}}_{{\bf a},\underline{{\bf b}}}\rangle} \xrightarrow[{N=\alpha n^2}]{n\gg1}
			\sum_{l=0}^{\left[ \frac{n}{2} \right]} \sum_{l_1=0}^{{\rm min}\left\{l,\left[\frac{n_1}{2}\right]\right\}}
			\sum_{k=0}^{{\rm min}\left\{l-l_1,n_1-2l_1\right\}} \\
 		\times C_{l,l_1,k} \left( \frac{-\epsilon}{\alpha} \right)^l n^{-2l} q_0^{l-k} q_1^k \,,
	\end{split}
\end{equation}
where the combinatorial factor for each tuple $(l,l_1,k)$ is given by
\begin{equation}
	\begin{split}
		C_{l,l_1,k} &=
			\underbrace{\left( \begin{matrix} n_1 \\ 2 l_1 \end{matrix} \right) (2 l_1 - 1)!!}_{\mbox{\small $l_1$ pairs in $\mathfrak{I}_1$}}
				\underbrace{\left( \begin{matrix} n_1 - 2 l_1 \\ k \end{matrix} \right) \left( \begin{matrix} n_0 \\ k \end{matrix} \right) k!}_{\mbox{\small $k$ pairs between $\mathfrak{I}_0$ and $\mathfrak{I}_1$}} \\
				&\quad \times \underbrace{\left( \begin{matrix} n_0 - k \\ 2l - 2 l_1 - 2 k \end{matrix} \right) (2l -2l_1 -2k -1)!!}_{\mbox{\small remaining pairs in $\mathfrak{I}_0$}} \\
		&= \frac{1}{2^{l-k} l_1! (l-l_1)!} \left( \begin{matrix} l - l_1 \\ k \end{matrix} \right) \\
		&\quad \times \underbrace{\left[ \frac{n_1!}{(n_1-2l_1-k)!} \frac{n_0!}{(n_0-2l+2l_1+k)!} \right]}_{\mbox{\small $\xrightarrow{n\gg1} n_1^{2l_1+k} n_0^{2l-2l_1-k}$}} \,.
	\end{split}
\end{equation}
The limiting value of the expression in square brackets is valid for any fixed $l,l_1$ and $k$. 
After expressing $n_0$ and $n_1$ by their (finite) fractions of $n$ the addends of~(\ref{eq:Pinfty1}) become independent of $n$ (due to the scaling $N=\alpha n^2$).
This is the point where it becomes evident that a scaling of $N$ with a power of $n$ different from $\eta = 2$ would lead to trivial results corresponding to Eq.~(\ref{eq:finII}) of the main text (for $\eta > 1$) also in the present case where $\tau_d \neq 0$ and $z_{ij} \neq 0$ for some $i,j$. Simplifying the upper limits of the sums for $n \rightarrow \infty$ yields
\begin{equation}
% 	\label{eq:QSuppFin}
	\begin{split}
		&\frac{\langle P^{(\epsilon)}_{{\bf a},\underline{{\bf b}}}\rangle}{\langle P^{{\rm (cl)}}_{{\bf a},\underline{{\bf b}}}\rangle} \xrightarrow[{N=\alpha n^2}]{n\gg1}
			\sum_{l=0}^\infty \sum_{l_1=0}^l \frac{1}{l!} \left( \frac{-\epsilon}{2 \alpha} \right)^l
			\left( \begin{matrix} l \\ l_1 \end{matrix} \right) \left(q_0 y_1^2\right)^{l_1} y_0^{l-l_1} \\
		&\qquad \qquad \qquad \times \sum_{k=0}^{l-l_1}  \left( \begin{matrix} l - l_1 \\ k \end{matrix} \right) (q_0 y_0)^{l-l_1-k} (2 q_1 y_1)^k \\
		&\;\;= \sum_{l=0}^\infty \frac{1}{l!} \left( \frac{-\epsilon}{2 \alpha} \right)^l
			\sum_{l_1=0}^l \left( \begin{matrix} l \\ l_1 \end{matrix} \right) \left(q_0 y_1^2\right)^{l_1} \left( 2 q_1 y_0 y_1 + q_0 y_0^2\right)^{l-l_1}\\
		&\;\;= \sum_{l=0}^\infty \frac{1}{l!} \left( \frac{-\epsilon}{2 \alpha} \right)^l \left(q_0 \left( y_0^2 + y_1^2 \right) + 2 q_1 y_0 y_1\right)^l%\\
% 		&\;\;= \exp \left[ - \frac{\epsilon}{4 \alpha} \left( {\cal Q}^{(2)}(0) \left(1+x^2\right) + {\cal Q}^{(2)}(z_1) \left(1-x^2\right) \right) \right] \,,
	\end{split} \nonumber
\end{equation}
\pagebreak
\begin{equation}
\label{eq:QSuppFin}
	\begin{split}
		&\;\;= \exp \left[ - \frac{\epsilon}{4 \alpha} \left( {\cal Q}^{(2)}(0) \left(1+x^2\right) + {\cal Q}^{(2)}(z_1) \left(1-x^2\right) \right) \right] \,,
	\end{split}
\end{equation}
where in the last step we rewrote $q_0$ and $q_1$ in terms of overlap functions and introduced the imbalance parameter $x$ related to $y_0$ and $y_1$ by
\begin{equation}
\label{eq:QSuppFinII}
	\begin{split}
		y_0 &= \frac{1}{2}(1-x) \,, \\
		y_1 &= \frac{1}{2}(1+x) \,.
	\end{split}
\end{equation}
Equation~(\ref{eq:QSuppFin}) is plotted and analysed in Fig.~\ref{fig:SP1} of the main text and corresponds to the result Eq.~(\ref{eq:finI}) there.
\vspace*{1.5cm}

%\JK{Make the starting number in the references match (it is the number in the following latex command)}

\makeatletter
\apptocmd{\thebibliography}{\global\c@NAT@ctr 53 \relax}{}{}
\makeatother


\begin{thebibliography}{99}
\bibitem{new} Due to the formal analogy with photonic systems, in the limit of structureless cavities ($\tau_{d}\to 0$), our mesoscopic BS can be also used to implement tasks like simulating vibronic spectra in molecules \cite{new1} and generating massive path entanglement for metrology applications as in \cite{new2}.
\end{thebibliography}
\end{document}